\documentclass[nopacs,nokeys]{revtex4}%
\usepackage{amsfonts}
\usepackage{amsmath}
\usepackage{amssymb}
\usepackage{bbm}
\usepackage[utf8]{inputenc}
\usepackage{tensor}
\usepackage{slashed}
\usepackage[centertableaux ]{ytableau}
\usepackage{hyperref}
\usepackage{natbib}
\usepackage{enumerate}
\usepackage{braket,mleftright}
\usepackage{graphicx}

\usepackage{tcolorbox}
\usepackage{color}
 \definecolor{jblue}  {RGB}{20,50,100}
  \definecolor{npurple}  {RGB} {153, 51, 204}
  \definecolor{wred}   {RGB}{217,0,56}
  \definecolor{white}   {RGB}{255,255,255}
  \definecolor{korange}   {RGB}{235, 80,  43}
  \definecolor{korange2}   {RGB}{245, 100,  63}
  \definecolor{kyelloworange}   {RGB}{255, 210,  110}
  \definecolor{kyelloworange2}   {RGB}{240, 170,  90}
  \definecolor{kred}   {RGB}{204,  102, 153}
  \definecolor{kpurple}   {RGB}{153,  61, 190}
  \definecolor{kpurplelight}   {RGB}{213,  161, 230}
 \definecolor{nblue}   {RGB}{200,  230, 250}
\definecolor{mycolor}{RGB}{5,61,245}

\usepackage{tikz}
\usetikzlibrary{trees}
\usetikzlibrary{snakes}
\usetikzlibrary{positioning,decorations.pathmorphing}
\usetikzlibrary{decorations.markings}
\usepackage{tikz,ifthen}
\tikzset{beamerprimary/.style={structure.fg, thick}}
\tikzset{beamersecondary/.style={structure.bg, thick}}
\tikzset{ boson/.style={decorate, decoration={snake}},
     gauge/.style={decorate,decoration={snake,post length=1mm}}  ,
     fermion/.style={postaction={decorate},decoration={markings,mark=at position .55 with {\arrow{>}}}},
    fermionloop/.style={postaction={decorate},
        decoration={markings,mark=at position .25 with {\arrow{<}}}}, 
    gluon/.style={decorate, 
        decoration={coil,amplitude=4pt, segment length=5pt}},
    scalar/.style={dotted,postaction={decorate},decoration={markings,mark=at position .75 with {\arrow{>}}}},
    graviton/.style={double},
    dm/.style={double,postaction={decorate},decoration={markings,mark=at position .55 with {\arrow{>}}}},
}

\begin{document}

\title{Supersymmetric Expansion Algorithm and complete analytical solution for the Hulth\'en and anharmonic potentials}
 \author{ M. Napsuciale$^{(1)}$, S. Rodr\'{\i}guez$^{(2)}$, M. Kirchbach$^{(3)}$ }
\address{$^{(1)}$Departamento de F\'{i}sica, Universidad de Guanajuato, Lomas del Campestre 103, Fraccionamiento
Lomas del Campestre, Le\'on, Guanajuato, 37150, M\'exico.}
\address{$^{(2)}$Facultad de Ciencias F\'isico-Matem\'aticas,
  Universidad Aut\'onoma de Coahuila, Edificio A, Unidad
  Camporredondo, Saltillo, Coahuila, 25000, M\'exico,} 
\address{$^{(3)}$Instituto de F\'{i}sica,
  Universidad Aut\'onoma de San Luis Potos\'{i}, Avenida Parque de Chapultepec 1570, San Luis Potos\'{i}, 78295, M\'exico.} 

\begin{abstract}
An algorithm for providing analytical solutions to Schr\"{o}dinger's  equation with non-exactly solvable potentials is elaborated. 
It represents a symbiosis between the logarithmic expansion method and the techniques of the superymmetric quantum mechanics 
as extended toward non shape invariant potentials. The complete solution to a given Hamiltonian $H_{0}$ is obtained from 
 the nodeless states of the Hamiltonian $H_{0}$ and of a set of supersymmetric partners $H_{1}, H_{2},..., H_{r}$. 
The nodeless states (dubbed ``edge" states) are unique and in general can be ground or excited states. They  are solved using 
the logarithmic expansion which yields an infinite systems of coupled first order hierarchical differential equations, converted later 
into algebraic equations with recurrence relations which can be solved order by order.  We formulate the aforementioned scheme, 
termed to as ``Supersymmetric Expansion Algorithm''   step by step and apply it to obtain for the first time the complete analytical 
solutions of the three dimensional Hulth\'en--, and the one-dimensional  anharmonic oscillator potentials. 

\end{abstract}
\maketitle

\section{Introduction}
The Schr\"{o}dinger equation occupies a pole position in the non-relativistic quantum mechanics as 
the knowledge of its solutions is the key prerequisite  for carrying out all the analyses of the observable characteristics  of quantum 
systems such as  bound and continuous spectra, scattering cross sections, transition probabilities, electric and magnetic multipole 
moments etc. In its simplest version it describes  two-body systems, such as the H Atom, di-atomic molecules etc. which allow for its 
reduction to a one-body problem. Due to the central symmetry of the majority of physics problems, it is often written in spherical polar 
coordinates   and upon variable separation, the angular part is described by the spherical harmonics, while its radial part contains the 
potential. Among the potentials, those admitting exact solutions, i.e., solutions which can be cast in terms of elementary functions and 
termed to as Liouvillian \cite{SINGER1991251}, are of special interest as they often  allow for realistic predictions of  the physical 
properties of the systems under consideration, being the Coulomb problem of the H Atom the example of excellence. This part of 
quantum mechanics will be termed to as ``Liouvillian Solvable Quantum Mechanics''.

\vspace {0.23cm}
\noindent
{\it Liouvillian Solvable Quantum Mechanics}

\vspace{0.23cm}

\noindent
Any potential in the  Liouvillian Solvable Quantum Mechanics can be solved by subjecting the corresponding Schr\"{o}dinger equation 
to a point-canonical transformation toward an exactly solvable second order differential equation of the type of the hyper-geometric-, 
Heun-, or other such equations \cite{R_De:1992} (see also \cite{Stevenson:1941} for earlier use of the method although not under the 
name of point-canonical transformation). As a rule, the solutions are exponentials times polynomials and their degrees, i.e. the numbers of 
nodes in the wave function constructed from them, are used as  quantum numbers in the characterizations of the excited states. In 
Schr\"odinger's work \cite{Schrodinger:1940xj}, another scheme has been suggested for finding the solutions of exactly solvable potentials. 
The scheme is based on factorization of the Hamiltonian by two conjugate operators, the definition of the ground state as a function 
annihilated by one of them, and generation of the excited states by the repeated action on the ground state by
 the other operator. Schr\"odinger's scheme  culminated in the work by Infeld and Hull \cite{Infeld:1951mw}  who extended it to include  
 a variety of  potentials and thus established the factorization method in quantum mechanics.

\vspace{0.23cm}
\noindent
{\it  Liouvillian Solvable SUSYQM}

\vspace{0.23cm}

\noindent
In the beginning of the 80ies Witten developed the idea about super-symmetric 
partner Hamiltonians \cite{Witten:1981nf} which, in combination with the factorization method, lead to an efficient  algebraic scheme for 
calculations  of energies and eigenfunctions of all the known exactly solvable quantum mechanical potential problems. In the subsequent 
decade, the so called super-symmetric quantum mechanics (SUSYQM) experienced a period of extensive development and produced  
many remarkably results, among them the possibility of designing  hierarchic chains constituted by partner Hamiltonians in such a way, 
that the ground state of the partner Hamiltonian is of same energy as the first excited state of the parent one, while the 
wave functions of the eigenstates to the  two Hamiltonians are related by specific differential operators termed to as ``intertwining''--, 
or,  ``shift'' operators,  a construct of many intriguing and useful  properties.
Compared to these activities, the quantum mechanical problems involving non-exactly solvable potentials,  remained out of the main 
scope and the consensus has been that they have to be treated numerically  and solved approximately by perturbative-, asymptotic 
iteration-, WKB and similar  methods. In what follows we here term to this branch of quantum mechanics as 
``No Liouvillian Solvable Quantum Mechanics''.

\vspace{0.23cm}

\noindent
{\it No Liouvillian Solvable Quantum Mechanics} 

 \vspace{0.23cm}
The potentials characterizing  the No Liouvillian Quantum Mechanics are such that do not possess solutions which can be cast in terms 
of elementary functions, the simplest examples being polynomial potentials with arbitrary coefficients.
Indeed, according to a theorem by Plesset \cite{Plesset:1932}, no exact solutions in terms of finite number of elementary functions can 
be found if the potential is a Laurent polynomial with arbitrary coefficients.
In 1972 Hautot \cite{Hautot:1972} formulated conditions on the aforementioned coefficients which allowed to resolve the resulting  
polynomial potentials exactly.
This work has been continued by Magyari \cite{Magyari:1981kx}.  The potentials conditioned in this way could be tackled equally well 
by the techniques of the Liouvillian solvable super-symmetric quantum mechanics \cite{Bera2017} and  thus be incorporated into the latter.

\noindent
Regarding  unconditioned polynomial potentials, progress has been achieved by the so called $\delta$ expansion technique \cite{Bender:1988rq} 
which solves the Schr\"odinger equation for a potential of the type, $V=V_{0}+V_{p}$, for cases in which  the exact  solutions to $V_{0}$ are 
known. More specifically, first an artificial parameter $\delta$ is attached to $V_{p}$,  and then an expansion in this parameter is performed, 
which interpolates between the known potential $V_{0}$ for $\delta=0$ and the potential $V$ for $\delta=1$. The expansion yields an infinite 
chain of coupled {\it second order} differential equations which can be solved  order by order  recursively. 
This scheme is especially suited for polynomial potentials with arbitrary coefficients such as anharmonic ones, in which case 
the coupled equations become algebraic and can be solved exactly, a reason for which such solutions are referred to as analytical.  
In this way,  the no Liouvillian solutions can be obtained order by order with arbitrary accuracy. Examples can be found in 
\cite{Bera:2007}, \cite{Datta:2011zz}.
\noindent

In 1940, Bijl \cite{BIJL1940869} suggested to apply expansion techniques to the Schr\"odinger equation for the logarithm of the wave function, i.e. to 
represent $\Psi (x)$ as
\begin{equation}
\Psi (r)=e^{\xi (x)}.
\label{lgrt_psi}
\end{equation}
In so doing one finds the following equation satisfied by $\varphi(x)=\frac{d}{dx}\xi (x)=\frac{d}{dx}\ln \Psi(x)$,
\begin{equation}
\varphi^2(x) + \varphi'(x)=\frac{2m}{\hbar^{2}}(V(x) -E),
\label{rct}
\end{equation}
known as Riccati's equation. Occasionally, (\ref{rct}) will also be referred to as ``logarithmic form''  of Schr\"odinger's framework.  Then an 
expansions in terms of an artificial parameter $\lambda$ is performed
\begin{eqnarray}
\varphi (x)&=&\varphi_0(x) +\lambda \varphi _1(x)+\lambda_2\varphi_2(x) +...,\nonumber\\
E&=&E_0 +\lambda E_1 +\lambda E_2 +....
\label{AB_expns}
\end{eqnarray}
Back substituting the latter expansions  into (\ref{rct}) amounts to an infinite chain of coupled, now {\it first order} differential equations which 
can be solved order by order  recursively. This formalism acquired in \cite{Aharonov:1979gxt} the name of  Logarithmic Perturbation (also 
Logarithmic Expansion)  technique.  Similarly to the $\delta$ expansion technique, it again leads to solutions of arbitrary accurateness order 
by order. A difficulty of this method is the lack of well defined functions  at the nodes of the wave functions, as they occur  for excited states.  
Many authors from the late 70ies onward have updated and applied this method and found the energies and ground state wave functions  to 
several  anharmonic or exponential potentials (see \cite{Eletsky:1981fm} and references therein as a representative example).

\vspace{0.23cm}

\noindent
{\it No Liouvillian Solvable SUSYQM }
\vspace{0.23cm}

\noindent
Non exactly or quasi-exactly  solvable potentials play a crucial role in quantum mechanical problems. Among them such of wide spread refer 
to anharmonic potentials and  potentials containing centrifugal barriers which without the latter throughout might be exactly solvable. The 
methods in the literature to tackle such problems are in their majority approximative, often  based on perturbative or WKB approaches. Some 
of them have been handled within the frameworks of the expansion techniques highlighted above. 
In 1990 the $\delta$ expansion  technique has been incorporated by Cooper and Roy into the framework of the super-symmetric quantum 
mechanics in \cite{Cooper1990EF}. As a reminder, the SUSYQM framework allows to produce the energies and wave functions in the 
potential's spectra by means of intertwining, i.e. shift operators and on the basis of the knowledge of ground states alone. In this fashion, 
a new branch of the super-symmetric quantum mechanics, here termed to as ``No Liouvillian solvable SUSYQM'' arose in relation  to 
potentials with no-Liouvillian solutions, and in parallel to the Liouvillian Solvable  SUSYQM, i.e. to the one devoted to  potentials whose  
solutions can be written  in closed form.
Finally, in 2000 Lee demonstrated in \cite{LEE2000101} the equivalence of the logarithmic expansion technique and SUSYQM with an 
incorporated $\delta$ expansion of the super-potential. Applications of this scheme can be found in \cite{Dhatt:2011}, where polynomial 
potentials have been considered. \\

\noindent
A full flashed consideration of an exponential potential (Yukawa's, to be precise) in a new formalism for No-Liouvillian Solvable SUSYQM  
has been reported  in \cite{Napsuciale:2020ehf,Napsuciale:2021qtw}.
It is the goal of the present work to systematize and lift up  the method employed in \cite{Napsuciale:2020ehf,Napsuciale:2021qtw} towards  
an algorithm applicable to a wide range of potentials of physical interest. 

The layout of the work is as follows. 
In Section II we elaborate step by step on the new approach dubbed ``Supersymmetric Expansion Algorithm (SEA)'', and explain 
how it incorporates several already known advances in the field next to  new elements and allows us to solve some of the  long pending 
potential problems. As a second example (after Yukawa's potential) of the applicability of the SEA, in section III we use it to completely 
solve the three dimensional  Hulth\'en potential. The complete analytical solution for the anharmonic potential and a proposal for a new 
supersymmetric formalism for perturbation theory are given in Section IV. The paper closes with concise conclusions in Section V.

\section{Supersymmetric Expansion Algorithm}
The Schr\"odinger equation for a particle of  mass $m$ in a one-dimensional potential $V(z,\lambda_{i})$ is
\begin{equation}
\left[-\frac{\hbar^{2}}{2m} \frac{d^{2}}{dz^{2}}+V(z,\lambda_{i}) \right]\psi(z,\lambda_{i})=E(\lambda_{i})~\psi(z,\lambda_{i}).
\end{equation}
The potential $V(z,\lambda_{i})$ depends in general on parameters $\lambda_{i},\, i=0,1,2,...,r$, related to the strength of the involved 
interactions or physical properties of the system. Each of these parameters defines a length scale $a_{i}$, and we are free to  single out one, 
to be denoted as $a_{0}$, and associated with, say, the parameter $\lambda_{0}$, with the aim to write Schr\"odinger's equation in terms 
of the dimensionless variable, $x=z/a_{0}$, arriving at 
\begin{equation}
\left[-\frac{d^{2}}{dx^{2}} +v(x,\lambda_{i}) \right]u(x,\lambda_{i})=\epsilon(\lambda_{i})~u(x,\lambda_{i}),
\end{equation}
with the following shorthand notations for the dimensionless quantities
\begin{equation}
v(x,\lambda_{i})=\frac{2m a^{2}_{0}}{\hbar^{2}} V(a_{0}x,\lambda_{i}), \qquad u(x,\lambda_{i})=\psi(a_{0}x,\lambda_{i}),
\qquad \epsilon(\lambda_{i})=\frac{2m a^{2}_{0}}{\hbar^{2}} E(\lambda_{i}).
\end{equation}

A version of the Supersymmetric Expansion Algorithm (SEA) whose full systematic will be introduced in this work, was already employed 
to obtain for the first time a complete analytical solution to the Yukawa potential in \cite{Napsuciale:2020ehf,Napsuciale:2021qtw}. As it will be 
shown below, it can furthermore  be applied to solve a wide class of potentials relevant in several branches of physics and chemistry and 
whose solutions have been considered as problematic so far, a reason for which  we consider it worth of systematization.  For the sake of 
simplicity in the notations from here onward we will consider only two parameters, $\lambda_{0}$ and $\lambda_{1}$. The algorithm is 
summarized by  the following ten steps:

\begin{enumerate}

\item Use the typical scale $a_{0}$ dictated by the parameter $\lambda_{0}$ to write the Schr\"odinger equation in terms of the 
dimensionless variable $x=z/a_{0}$ and the associated  dimensionless quantities as
\begin{equation}
\left[-\frac{d}{dx^{2}} + v_{0}(x,\lambda)\right]u_{0}(x,\lambda)=\epsilon_{0}(\lambda)u_{0}(x,\lambda).
\label{SE0}
\end{equation}
Here we renamed to $\lambda$  the parameter $\lambda_{1}$  as it  emerges upon  the change of variable in (\ref{SE0}). 
As this will turn out to be the first step in an iteration the procedure, to be described below,  we attached a 
subscript "$0$" to all the  quantities appearing at this stage.
 Within  these notations, we are interested in the solutions to
\begin{equation}
H_{0}=-\frac{d^{2}}{dx^{2}}+v_{0}(x,\lambda).
\end{equation}

\item Cast Eq.(\ref{SE0}) in the form of a Riccati equation as 
\begin{equation}
W_{0}^{2}(x,\lambda) - W_{0}^{\prime}(x,\lambda) = v_{0}(x,\lambda)-\epsilon_{0}(\lambda),
\label{Riccati}
\end{equation}
where 
\begin{equation}
W_{0}(x,\lambda)\equiv -\frac{d}{dx}\ln u_{0}(x,\lambda),
\label{W0}
\end{equation}
is the superpotential.

\item Expand the potential in powers of $\lambda$ according to,
\begin{equation}
v_{0}(x,\lambda)=\sum_{k=0}^{\infty}v_{0k}(x) \lambda^{k},
\end{equation}
and expand accordingly  $W_{0}(x,\lambda)$  in(\ref{W0}) and the energy $\epsilon_{0}(\lambda)$ as 
 \begin{equation}
W_{0}(x,\lambda)=\sum^{\infty}_{k=0} w_{0k}(x)\lambda^{k}  , \qquad  
\epsilon_{0}(\lambda)=\sum^{\infty}_{k=0}  \varepsilon_{0k}\lambda^{k}.
\label{solW0}
\end{equation}

\item Re-design  the product of the two infinite series defining the non-linear term $W_{0}^{2}$ in (\ref{Riccati})  
as  a single infinite series according to,
\begin{equation}
W_{0}^{2}=\left(\sum^{\infty}_{m=0}w_{0m} \lambda^{m} \right)\left(\sum^{\infty}_{n=0} w_{0n}  \lambda^{n}\right)
=\sum^{\infty}_{m,n=0}w_{0m}w_{0n} \lambda^{m+n} =\sum^{\infty}_{k=0} C_{0k} \lambda^{k}.
\end{equation}
Here,  the  $C_{0k}$ coefficients are given by the following finite sums,
 \begin{equation}
C_{0k}=\sum_{m+n=k}w_{0m}w_{0n}= \left\{ 
\begin{array}{cc}
w^{2}_{00}, & k=0 \\
2w_{00}w_{0k} + B_{0k}, & k\ge 1 
\end{array}
\right. .
\end{equation}
Also note that
\begin{equation}
B_{0k}=\sum_{\substack{m+n=k \\ m,n\neq 0 }} w_{0m}w_{0n},
\end{equation}
does not involve $w_{0k}$.

\item Use these elements in the logarithmic formulation of  Schr\"odinger's equation in (\ref{Riccati}) to obtain the infinite set of coupled  first order differential equations
\begin{align}
w_{00}^{2} - w'_{00} &= v_{00}(x) - \varepsilon_{00},  \label{w00eq}\\
2w_{00}w_{0k} -  w^{\prime}_{0k}&= - B_{0k} + v_{0k} -  \varepsilon_{0k} , \qquad k=1,...\infty.
\label{w0keq}
\end{align}
Explicitly, we find
\begin{align}
k&=0:  & w_{00}^{2} - w'_{00} &= v_{00}(x) - \varepsilon_{00}, \\
k&=1:  & 2 w_{00} w_{01}- w'_{01} &=v_{01}(x)-\varepsilon_{01}, \\
k&=2:  & 2 w_{00} w_{02}  - w'_{02} &= -  w_{01}^{2}+v_{02}(x)-\varepsilon_{02},  \\
k&=3:  & 2 w_{00} w_{03}  - w'_{03} &= -2 w_{01} w_{02} + v_{03}(x)-\varepsilon_{03}, \\
k&=4:  & 2 w_{00} w_{04}  - w'_{04} &= -2 w_{01} w_{03} -2 w^{2}_{02}  + v_{04}(x)-\varepsilon_{04}, \\
...& ... &... &... \nonumber
\end{align}
This is a cascade of linear differential equations, where the $k$-th equation involves $w_{0k}$ next to the solutions for $k-1,k-2,...,1,0$. We can
solve it  starting from the equation for $k=0$.

\item For a wide class of potentials the coefficients $v_{0k}(x)$ are Laurent polynomials in $x$, and one finds
\begin{equation}
v_{0k}(x)=\sum_{\alpha=-b}^{t}v_{0k\alpha}x^{\alpha}.
\end{equation}
This expression includes both exponential and polynomial potentials, or any combination of them,  among which those of highest importance are 
the potentials by Yukawa and Hulth\'en, the anharmonic potentials, not to forget the  already solved Coulomb and harmonic oscillator potentials, 
in which  case the solutions in $x$ for $w_{0k}$ are polynomials,
\begin{equation}
w_{0k}(x)=\sum_{\alpha=\alpha_{min}}^{s(k)} w_{0k\alpha}x^{\alpha},
\label{w0ka}
\end{equation}
with  $w_{0k\alpha}$ being constant coefficients. The values of $\alpha_{min}$ and $s(k)$ depend on the potential and need to  be worked out
for each specific case. If we were able to solve the $k=0$ case in Eq.(\ref{w00eq}), then the remaining equations would become algebraic, 
and recurrence relations could be obtained for the coefficients $w_{0k\alpha}$ and $\varepsilon_{0k}$, obtaining order by order an analytical 
solution for the state $u_{0}(x,\lambda)$ and its eigenvalue $\epsilon_{0}(\lambda)$. 

In particular, if the potential $v_{0}(x,\lambda)$ is such that the solutions for $v_{0}(x,0)=v_{00}(x)$ are known, then the procedure outlined 
allows to  obtain  a solution for $v_{0}(x,\lambda)$. This category includes but do not restrict to screening potentials and potentials of the form 
$v(x,\lambda)=v_{00}(x) + \lambda v_{p}(x)$.  The solution following
from Eq. (\ref{W0}) reads 
\begin{equation}
u_{0}(x,\lambda)=N(\lambda)e^{-\int W_{0}(x,\lambda) dx },
\end{equation}
where $N(\lambda)$ is a normalization factor. Notice that by construction, {\emph{this solution has no nodes}}. This is an important point because 
we will show below that for some potentials relevant in physics and chemistry (e.g. screening potentials) this first solution corresponds to an excited state 
and in the past the main obstruction to the solution for excited states has been the assumed existence of nodes for all excited states.

The Hamiltonian $H_{0}$ can be factorized as
\begin{equation}
H_{0}=a^{\dagger}_{0}a_{0} + \epsilon_{0}(\lambda),
\end{equation}  
with the factorization operators $a_{0}$ and $a_{0}^{\dagger}$ being given by
\begin{equation}
a^{\dagger}_{0}= - \frac{d}{dx} + W_{0}(x,\lambda), \qquad  a_{0}=\frac{d}{dx} + W_{0}(x,\lambda),
\label{ladderop}
\end{equation}
while  the wave function  $u_{0}(x,\lambda)$ is annihilated by $a_{0}$ as
\begin{equation}
a_{0}u_{0}(x,\lambda)=0.
\end{equation}

\item The wave functions to the remaining states are then solved by the aid of
 techniques known from
the supersymmetric quantum mechanics. One begins with constructing  the first  supersymmetric partner $H_{1}$ to $H_0$ as
\begin{equation}
H_{1}=a_{0}a_{0}^{\dagger} + \epsilon_{0}(\lambda)= -\frac{d^{2}}{dx^{2}}+ v_{1}(x,\lambda).
\end{equation}  
The associated potentials are then related by
\begin{equation}
 v_{1}(x,\lambda)= v_{0}(x,\lambda) + 2 W_{0}^{\prime}(x,\lambda).
\end{equation}
The superpartners $H_{0}$ and $H_{1}$ share  same spectrum and 
have same number of states, except for one state, the ``ground" state to $H_0$. 
Indeed, suppose we  find $u_1(x)$ as an 
eigenstate to $H_{1}$, i.e. 
\begin{equation}
H_{1} u_{1}(x)=[a_{0}a_{0}^{\dagger} + \epsilon_{0}(\lambda) ] u_{1}(x)= \epsilon_{1} u_{1}(x).
\end{equation}
Then
\begin{equation}
[a_{0}^{\dagger}a_{0}+ \epsilon_{0}(\lambda)]a_{0}^{\dagger}u_{1}(x)=H_{0}(a_{0}^{\dagger}u_{1}(x))= \epsilon_{1} (a_{0}^{\dagger}u_{1}(x)),
\end{equation}
therefore, $\epsilon_{1}$ is another eigenvalue of the $H_0$ with eigenstate $u_{01}(x)=a_{0}^{\dagger}u_{1}(x)$. 

\item We apply the steps 2 to 6 to the partner Hamiltonian $H_{1}$ in order to obtain a solution to the Riccati equation 
for the corresponding super-potential $W_{1}(x,\lambda)$ , satisfying
\begin{equation}
W_{1}^{2}(x,\lambda) - W_{1}^{\prime}(x,\lambda)=v_{1}(x,\lambda)-\epsilon_{1}(\lambda).
\end{equation}
The solution so found yields the following {\emph{nodeless}} eigenstate  to $H_{1}$
\begin{equation}
u_{1}(x,\lambda)=N_{1}(\lambda)e^{-\int W_{1}(x,\lambda)dx}.
\end{equation}
It satisfies
\begin{equation}
H_{1} u_{1}(x,\lambda)=\epsilon_{1}(\lambda) u_{1}(x,\lambda),
\end{equation}
and yields, according to Step 7, a further solution to $H_{0}$ according to,
\begin{equation}
H_{0}\left[a^{\dagger}_{0}u_{1}(x,\lambda)\right] = \epsilon_{1}(\lambda)\left[a_{0}^{\dagger}u_{1}(x,\lambda) \right].
\end{equation}
This new solution to $H_{0}$ has nodes appearing due to the action of the operator $a_{0}^{\dagger}$ on $u_{1}(x,\lambda)$.

The Hamiltonian $H_{1}$ can now be factorized by itsef  as 
\begin{equation}
H_{1}=a^{\dagger}_{1}a_{1} + \epsilon_{1}(\lambda),
\end{equation}  
where
\begin{equation}
a^{\dagger}_{1}= - \frac{d}{dx} + W_{1}(x,\lambda), \qquad  a_{1}^{\dagger}=\frac{d}{dx} + W_{1}(x,\lambda),
\end{equation}
while  the state $u_{1}(x,\lambda)$ satisfies
\begin{equation}
a_{1}u_{1}(x,\lambda)=0.
\end{equation}
\item A further  solution to $H_{0}$ can be obtained upon constructing the  super-partner to $H_{1}$, denoted by $H_{2}$, and defined as
\begin{equation}
H_{2}=a_{1}a_{1}^{\dagger} + \epsilon_{1}(\lambda)= -\frac{d^{2}}{dx^{2}}+ v_{2}(x,\lambda).
\end{equation}  
The potentials are related as
\begin{equation}
 v_{2}(x,\lambda)= v_{1}(x,\lambda) + 2 W_{1}^{\prime}(x,\lambda).
\end{equation}
The corresponding superpotential satisfies
\begin{equation}
W_{2}^{2}(x,\lambda) - W_{2}^{\prime}(x,\lambda)=v_{2}(x,\lambda)-\epsilon_{2}(\lambda),
\end{equation}
which can be solved applying steps 2 to 6 now to $H_{2}$ to obtain the {\emph{nodeless}} solution
\begin{equation}
u_{2}(x,\lambda)=N_{2}(\lambda)e^{-\int W_{2}(x,\lambda)dx}.
\end{equation} 
The corresponding eigenvalue $\epsilon_{2}(\lambda)$ is shared by both
$H_{1}$ and $H_{0}$, being the related eigenstates 
$a^{\dagger}_{1}u_{2}(x,\lambda)$, and $a^{\dagger}_{0}a^{\dagger}_{1}u_{2}(x,\lambda)$, respectively. These eigenstates have nodes 
generated by the action of $a^{\dagger}_{1}$ and $a^{\dagger}_{0}a^{\dagger}_{1}$ on the nodeless state $u_{2}(x,\lambda)$ respectively.
\item We continue this process until we completely solve the original Hamiltonian $H_{0}$. For each Hamiltonian $H_{r}, \, r=0,1,2...$, in the previous steps, 
the set of equations 
\begin{align}
w^{2}_{r0} -  w^{\prime}_{rk}&=  v_{r0} -  \varepsilon_{r0},  \label{wr0eq} \\ 
2w_{r0}w_{rk} -  w^{\prime}_{rk}&= - B_{rk} + v_{rk} -  \varepsilon_{rk} , \qquad k=1,2,...\infty, \label{wrkeq}
\end{align}
becomes algebraic with a hierarchical structure, and recurrence relations can be obtained for the coefficients $w_{rk\alpha}$ and $\varepsilon_{rk}$, thus
completely solving the problem. 

\end{enumerate}

We remark that only few of the eigenvalues and eigenstates of the Hamiltonian we are interested in, $H_{0}$, are directly obtained by applying to it the
logarithmic formalism and the expansion of the potential, in a way that yields a nodeless solution. The remaining eigenstates and eigenvalues of $H_{0}$ are 
indirectly obtained solving first the supersymmetric partners $H_{r}$ (which yields the common eigenvalue) and applying the factorization operators on these 
(also nodeless) solutions. In each step increasing $r$, we obtain a nodeless solution of $H_{r}$ which upon the action of the factorization operators 
yields an eigenstate of $H_{0}$ in the form of a polynomial times the nodeless $H_{r}$ eigenstate. It is clear that the solution is build up on the 
initial nodeless solutions of $H_{0}$ and the superpartners $H_{r}$. It is a wide spread belief that for any Hamiltonian the only nodeless state is the 
ground state. We find that this not generally true, and there exist nodeless excited states which in our formalism are instrumental in solving the problem. 
For this reason, we single out these states and dub them {\it{edge}} states. The name comes from the extremal values that these states have for the 
angular momentum quantum number in the case of screening potential, but in general edge states can be defined as nodeless (excited or ground) states.
 
The Supersymmetric Expansion Algorithm combines several achievements  in the literature, to be listed below, and also adds some
new elements to yield a quite a robust framework. Steps 2-5 resemble the formalism of the Logarithmic Perturbation 
Theory (LPT) developed for the calculation of perturbative corrections to the {\emph{ground states}} of several potentials in an 
formalism alternative to the Rayleigh-Schr\"odinger (RS) theory. In Refs. \cite{Aharonov:1979gxt}  \cite{Dolgov:1978bc} relations  analogous to 
Eqs.(\ref{w00eq},\ref{w0keq}) were obtained for perturbed ground states, which were  solved using as integrating factor unperturbed ground states. 
The corrections to arbitrary order $k$ were obtained in the form of integrals over the whole space involving the ground state probability function, 
the perturbation potential, and the lowest order corrections $k-1,k-2...1,0$. Correction to excited states, expanded in powers of the perturbation 
parameter, were calculated by writing them as a products of the  ground state, cast into an exponential form,  and a polynomial, containing the 
nodes of the perturbed states. 
 At this point the calculation becomes cumbersome and its efficiency with respect to the RS formalism gets lost. There have been several attempts 
 in the literature to avoid the problem with handling  the nodes of  the excited states, a difficulty  which limits applications of the logarithmic 
 expansion method (see \cite{Bandyopadhyay:2002} and references therein for  alternatives).

 The same idea was applied to solve for the ground state of the screened coulomb potential in \cite{Eletsky:1981fm}  \cite{Vainberg:1981}
 where it could be  shown that  Eqs.(\ref{w00eq},\ref{w0keq}) have polynomial solutions. The  logarithmic perturbation theory (LPT)  could be 
applied to  the aforementioned ground state only because it was nodeless, and $W_{0}$ could well be  defined in this case. 
In the present work we will  show below how excited states with nodes, associated with a principle quantum number $n$,  
can be build up from the only nodeless excited state (the ``edge'' state) through the action on it by the factorization operators of supersymmetry. 
In this fashion, the logarithmic expansion method, once supplemented  by the technique of the supersymmetric quantum mechanics, 
allows to avoid the old problem of nodes in the excited states. A  similar option  has already been considered in \cite{LEE2000101}, where a comparison 
of LPT with  SUSY quantum mechanics, combined with  a $\delta$-expansion {\it a la }\cite{Bender:1988rq} of the superpotential  has been undertaken. 
 
 Another related development is the use of SUSY for a restricted class of potentials which have the property dubbed "shape invariance" 
 \cite{Gendenshtein:1984vs}. In this 
 case, the super-partner $H_{r}$ generated in steps 7 to 10 belongs to the same family of Hamiltonians differing only by a function of 
 the parameter $\lambda$. The harmonic oscillator and Coulomb potentials belong to this category. In this case it is possible to obtain 
 straightforwardly the whole set of eigenvalues and eigenfunctions in closed forms. Shape invariance is a property of supersymmetric 
 Hamiltonians which are factorizable as defined in Ref. \cite{Infeld:1951mw}. 
In the SEA formalism, $H_{r},\,r=1,2..$ are a chain of SUSY partners although shape invariance is not at all  required, thus extending the 
field for applications of  the techniques of supersymmetric quantum mechanics.

Closely related to the present work is the one in  \cite{Dhatt:2011},  where a formalism for perturbation theory is proposed  for potentials of the form 
$v(x)=v_{0}(x) +\lambda v_{p}(x)$. Polynomial solutions are assumed and the ground state is also solved using the expansion for the superpotential 
in the ground state. The problems of the nodes is avoided by constructing 
supersymmetric partner potentials which are directly and recursively written starting with the super-potential for the ground state. In this manner, 
the set of potentials $V_{r}$ are constructed and the excited states are obtained acting successively with intertwining  operators on the ground 
states of  $H_{r}$'s. The super-partners are solved using similar expansions for their own ground states. There are, however, several 
striking differences with our formalism. Firstly, the scheme in \cite{Dhatt:2011} is designed for a perturbative calculation. This is not the case of our 
SEA framework, although it can also be used for such purposes. Secondly, the assumption of polynomial solutions for the super-potentials works 
only for polynomial perturbations. Finally, but even more important, the aforementioned scheme  is designed to start the sequence by the ground state. 
This is convenient for unperturbed potentials, as is the harmonic oscillator, used in the examples worked out there. In our SEA framework, we show 
instead that in general it is possible to start the iterative procedure with  a properly chosen excited state (the edge state), 
which enlarges the applicability of our method.

\section{Supersymmetric Expansion Algorithm at work: Hulth\'en potential}

The Supersymmetric Expansion Algorithm was employed in \cite{Napsuciale:2020ehf,Napsuciale:2021qtw} to obtain a complete analytical 
solution for the Yukawa potential \footnote{The convention for $a$ and $a^{\dagger}$ operators used in 
\cite{Napsuciale:2020ehf,Napsuciale:2021qtw} is exchanged here to match the more common convention in the literature for these operators.}. 
In this section we will apply  it to completely solve the Hulth\'en potential, giving more  details on the  derivation of the algebraic solution via  the recurrence 
relations, which have not  not been presented in \cite{Napsuciale:2020ehf,Napsuciale:2021qtw}.

Proposed originally to describe nuclear interactions, the Hulth\'en potential \cite{Hulthen:1942} finds applications in atomic-, 
\cite{Tietz:1961,Lai:1980}, solid-state-- \cite{Berezin:1986}, and 
chemical physics \cite{Pyyokko:1975}. Exact solutions have been reported only 
for the $\ell=0$ states  \cite{Lam:1971,Flugge:1999}. 
Approximate solutions for the $\ell\neq 0$ states have been obtained using a variety of methods \cite{Lai:1980, Patil:1984, Roy:1987,Gonul:2000kix}.
Results on the numerical solutions as well as a comparison with previous approximate results can be found in Ref. \cite{Varshni:1990zz}.
To the best of our knowledge, the  complete analytical solutions to Schr\"{o}dinger's equation with this potential are still missing and it is our goal 
here to fill this gap upon applying the SEA framework described in the previous section.

The Hulth\'en potential is given by
\begin{equation}
V(r,D)=  \frac{\alpha_{h}\hbar c}{D}\frac{e^{-r/D}}{1-e^{-r/D}}.
\end{equation}

The typical distance scale for the Hulth\'en potential associated to the coupling $\alpha_{h}$ is given by $a_{h}=\hbar/mc\alpha_{h}$. 
Using this scale, the dimensionless potential can be written in either one of the following three forms,
\begin{equation} 
v_{h}(x,\lambda)= -\frac{2\lambda e^{-\lambda x}}{1-e^{-\lambda x}} = -\frac{2\lambda}{e^{\lambda x}-1} = \lambda\left(1-\coth\frac{\lambda x}{2}\right),
\end{equation}
where $\lambda=a_{h}/D$. We need to solve the Schr\"{o}dinger equation (\ref{SE0}) for the effective dimensionless potential and energy given by
\begin{equation}
v_{0}(x,\lambda)\equiv \frac{\ell(\ell+1)}{x^{2}} -\frac{2\lambda}{e^{\lambda x}-1}, \qquad \epsilon_{0}(\lambda)=\frac{2E(\lambda)}{mc^{2}\alpha^{2}_{h}},
\end{equation}
The potential allows for the following expansion,
\begin{align}
v_{0}(x,\lambda)&= \frac{\ell(\ell+1)}{x^{2}} - 2\sum_{k=0}^{\infty} \frac{B_{k}^{-} x^{k-1}\lambda^{k}}{k!} 
\end{align}
with the Bernoulli's numbers being,
\begin{equation}
B^{-}_{k}=\sum_{n=0}^{k}\sum_{m=0}^{n} (-1)^{m}\begin{pmatrix}n \\ m\end{pmatrix} \frac{m^{k}}{n+1}.
\end{equation}
This expression  yields the following expansion coefficients, 
\begin{align}
v_{0k}(x) =\left \{
\begin{array}{r}
\frac{\ell(\ell+1)}{x^{2}}-\frac{2}{x} \quad \textit{for} \quad k=0,\\ 
\frac{h_{k}x^{k-1}}{k!}\quad \textit{for} \quad k\ge 1 ,
  \end{array}
  \right.
\end{align}
where
\begin{equation}
h_{k}= - 2B^{-}_{k}.
\end{equation}
The set of equations (\ref{w00eq},\ref{w0keq}) in this case reads
\begin{align}
w_{00}^{2} - w'_{00} &= \frac{\ell(\ell+1)}{x^{2}}-\frac{2}{x} - \varepsilon_{00},  \\
2w_{00} w_{0k} - w'_{0k} &= -B_{0k} -\frac{2B^{-}_{k}x^{k-1}}{k!} -\varepsilon_{0k}.  \label{w0kheq} 
\end{align}
 The solutions up 
to $k=5$ are given by
\begin{align}
w_{00}(x)&= \frac{1}{b}-\frac{b}{x},  & \varepsilon_{00}&=-\frac{1}{b^{2}}, \label{w00h}\\
w_{01}(x)&= 0,  &\varepsilon_{01}&=1,  \label{w01h}\\
w_{02}(x)&= -\frac{b}{12}x, & \varepsilon_{02}&=-\frac{1}{12}b(2b+1), \label{w02h} \\
w_{03}(x)&= 0, & \varepsilon_{03}&=0, \label{w03h}\\
w_{04}(x)&= -\frac{b^{3}(b-1)(b+1)}{480} x  -\frac{b^{2}(b-1)}{480} x^{2} +\frac{b}{720} x^{3} , 
 &\varepsilon_{04}&=-\frac{1}{480}b^{3}(b-1)(b+1)(2b+1) ,  \label{w04h}\\
w_{05}(x)&= 0 , & \varepsilon_{05}&=0, \label{w05h} 
\end{align}
where $b=\ell+1$. We can see here that for $k\ge 1$ the solutions for the expansion coefficients of the superpotential are of the form
\begin{equation}
w_{0k}(x)=\sum_{\alpha=1}^{k-1} w_{0k\alpha}x^{\alpha}.
\label{wokah}
\end{equation}
However, we obtain more information
from the solutions of the first few terms which is crucial for the formalism. Indeed, the leading term in the Hulth\'en potential corresponds to the Coulomb 
potential which has energies $\epsilon^{c}=-1/n^{2}$, thus solutions to the Hulth\'en potential can also be labelled by a principal quantum number $n$ in
addition to the angular momentum $\ell$ and our formalism at this point yields the solution for $\ell=n-1$. The solution to the Riccati 
equation in this step reads
\begin{equation}
u_{0}(x,\lambda)=N_{0}e^{-\int W_{0}(x,\lambda)}dx\equiv \phi^{0}_{n,n-1}(x,\lambda),
\end{equation}
where $n$ is arbitrary and we have the first six terms of $W_{0}(x,\lambda)$ in Eq. (\ref{solW0}). At this point, it is convenient to separate the
$\lambda$-independent contribution and to write
\begin{equation}
W_{0}(x,\lambda)=\frac{1}{n}-\frac{n}{x} + w_{0}(x,\lambda),
\label{W0h}
\end{equation}
where
\begin{equation}
w_{0}(x,\lambda)=\sum_{k=1}^{\infty}\lambda^{k} w_{0k} (x).
\label{woh}
\end{equation}
The  solution can be rewritten as
\begin{equation}
u_{0}(x,\lambda)=N_{0}(\lambda) x^{n}e^{-\frac{x}{n}} e^{-G_{0}(x,\lambda)}= \phi^{0}_{n,n-1}(x,\lambda),
\label{edge}
\end{equation}
with
\begin{equation}
G_{0}(x,\lambda)=\int w_{0}(x,\lambda)dx.
\end{equation}
Notice that for $n\neq 1$ the solution in Eq.(\ref{edge}) describes  an excited state. This sounds strange on the light of the widely 
spread belief that all the excited states have nodes and the logarithmic formulation cannot be applied to them because the logarithm 
is ill-defined at the nodes. We can see explicitly in Eq. (\ref{edge}) however, that the state $ \phi^{0}_{n,n-1}(x,\lambda)$ has multiple 
zeroes at $x=0$ which is an end point, thus strictly speaking these states have no nodes and can be obtained using the logarithmic 
formalism. 
\begin{quote}
For a given $n$, the $\ell=n-1$ state is the only state with this property  and we  single it out, dubbing  it 
{\emph{edge}} state within 
 the $n$ level, this  because it corresponds to the  maximal $\ell$ 
value. 
\end{quote}
There is an edge state for every $n$ level and all the $ \phi^{0}_{n,n-1}(x,\lambda)$ states are given by Eq.(\ref{edge}) with the 
different values of $n$ distinguishing them. Notice also that there exist edge states for every potential $v_{0k}$ in the calculation 
at a given order $\lambda^{k}$. In particular, we have nodeless excited states for every $n$ level  in the Coulomb case which 
corresponds to the calculation to order $\lambda^{0}$ with $G_{0}(x,\lambda)=0$.

The edge states will be the starting point to solve the remaining $ \phi^{0}_{n,\ell}(x,\lambda)$ states with $\ell\neq n-1$. Since $n$ is arbitrary 
the procedure yields the complete solution for the Hulth\'en potential. With this aim, we insert the expansion (\ref{wokah}) in 
Eqs. (\ref{w0kheq}), for $k\ge 1$ to get
\begin{equation}
 \frac{2}{b}\sum_{\alpha=1}^{k-1}w_{0k\alpha}x^{\alpha} 
- \sum_{\alpha=0}^{k-2} (2b+\alpha+1)w_{0k(\alpha+1)} x^{\alpha} + \sum_{\alpha=2}^{k-2}B_{0k\alpha}x^{\alpha} 
- \frac{y_{k} }{k!}x^{k-1} + \varepsilon_{0k}=0,
\label{Bw0k}
\end{equation}
where
\begin{equation}
B_{0k\alpha}=\sum_{\substack{m+n=k \\ m,n\neq 0 }}\sum_{\beta+\gamma=\alpha}w_{0m\beta}w_{0n\gamma},
\end{equation}
which yields the following recurrence relations:
\begin{align}
w_{0k(k-1)}&=\frac{b}{2} \frac{h_{k}}{k!},   & & \label{w0kkm1y}\\
w_{0k\alpha}&=\frac{b}{2}\left[ (2b+\alpha+1)w_{0k(\alpha+1)}-B_{0k\alpha}\right], & \alpha&=k-2, k-1, ...2, \label{w0kay}\\
w_{0k1}&=b(b+1)w_{0k2}, & & \label{w0k1y}\\
\varepsilon_{0k}&=(2b+1)w_{0k1} + 2 \delta_{k1}. & &  \label{e0ky}
\end{align}
The coefficients $B_{0k\alpha}$ include products of $w_{0m\alpha}$ with $m=k-1,k-1,...1$, which in the $k$-step are already known, 
thus, these recurrence relations completely solve the problem for the edge state $u_{0}=\phi^{0}_{n,n-1}$  in Eq.(\ref{edge}) 
with
\begin{equation}
G_{0}(x,\lambda)=\int w_{0}(x,\lambda)dx = \sum_{k=1}^{\infty}\lambda^{k}\left( \sum_{\alpha=1}^{k-1}w_{0k\alpha} \frac{x^{\alpha+1}}{\alpha+1} \right).
\end{equation}

Now we continue with the steps 7-8 in the SEA. The first partner potential is given by
\begin{align}
v_{1}(x,\lambda)&=v_{0}(x,\lambda)+2 W^{\prime}_{0}(x,\lambda)=
 \frac{(\ell+1)(\ell+2)}{x^{2}}-\frac{2}{x} + \sum_{k=1}^{\infty}(v_{0k}+2w_{0k}^{\prime}) \lambda^{k}=\sum_{k=0}^{\infty}v_{1k}(x)\lambda^{k}.
 \label{v1h}
\end{align}
The coefficients  for $k\ge 1$ 
\begin{equation}
v_{1k}(x)=v_{0k}+2w_{0k}^{\prime}=\frac{h_{k}}{k!}x^{k-1} + 2 \sum_{\alpha=1}^{k-1}\alpha w_{0k\alpha} x^{\alpha-1},
\end{equation}
are polynomials in $x$. We solve this potential in a similar way as $H_{0}$. Expanding the superpotential $w_{1}(x,\lambda$) and the 
energy $\epsilon_{1}(\lambda)$, the Riccati equation for $H_{1}$
yields for $k=0$
\begin{equation}
w_{10}(x)=\frac{1}{b+1} - \frac{b+1}{x}, \qquad \varepsilon_{10}=-\frac{1}{(b+1)^{2}},  
\end{equation} 
while for $k\ge1$ produces the cascade of equations
 \begin{align}
2\left(\frac{1}{b+1}-\frac{b+1}{x}\right)w_{1k} - w^{\prime}_{1k} =- \sum_{\substack{m+n=k\\ m,n\neq 0}} w_{1m} w_{1n}  +
\frac{h_{k}}{k!}x^{k-1} + 2 \sum_{\alpha=1}^{k-1}\alpha w_{0k\alpha} x^{\alpha-1}-\varepsilon_{1k}.  \label{w1ky}
\end{align}
Expanding now
\begin{equation}
w_{1k}(x)=\sum_{\alpha=1}^{k-1} w_{1k\alpha} x^{\alpha},
\end{equation} 
and inserting the expansion in Eq.(\ref{w1ky}) amounts to
\begin{equation}
 \frac{2}{b+1}\sum_{\alpha=1}^{k-1}w_{1k\alpha}x^{\alpha} 
- \sum_{\alpha=0}^{k-2} \left[ (2(b+1)+\alpha+1)w_{1k(\alpha+1)} + 2(\alpha+1) w_{0k(\alpha+1)} \right]x^{\alpha} 
+\sum_{\alpha=2}^{k-2}B_{1k\alpha}x^{\alpha} 
- \frac{h_{k}}{k!}x^{k-1} + \varepsilon_{1k}=0,
\end{equation}
where
\begin{equation}
B_{1k\alpha}=\sum_{\substack{m+n=k\\ m,n\neq 0}}\sum_{\beta+\gamma=\alpha}w_{1m\beta}w_{1n\gamma}.
\end{equation}
We obtain the following recurrence relations,
\begin{align}
w_{1k(k-1)}&=\frac{b+1}{2}\frac{h_{k}}{k!}   & & \label{w1kkm1y} \\
w_{1k\alpha}&=\frac{b+1}{2}\left[ (2(b+1)+\alpha+1)w_{1k(\alpha+1)}+2(\alpha+1) w_{0k(\alpha+1)} -B_{1k\alpha}\right], & \alpha&=k-2, k-1, ...2. \label{w1kay}\\
w_{1k1}&=\frac{b+1}{2}\left[ 2(b+2)w_{1k2}+4 w_{0k2}\right], & &  \label{w1k1y} \\
\varepsilon_{1k}&=(2b+3)w_{1k1} + 2w_{0k1} + 2 \delta_{k1}. & & \label{e1ky} 
\end{align}

These recurrence relations completely solve the problem for $H_{1}$. Notice that $H_{1}$ is different from $H_{0}$ but it is also a screening Hamiltonian 
in the sense that it reduces to the Coulomb Hamiltonian in the $\lambda \to 0$ limit. This allows us to identify the eigenstate, here denoted by  
$u_{1}(x,\lambda)\equiv \phi^{1}_{n,n-2}(x,\lambda)$, of the leading term of
$H_{1}$ (though not of $H_{0}$) as a $\ell=(n-2)$-state.
  Explicitly, it is given by
\begin{equation}
u_{1}(x,\lambda)= x^{n}e^{-\frac{x}{n}} e^{-G_{1}(x,\lambda)}=\phi^{1}_{n,n-2}(x,\lambda),
\end{equation}
where we from here onward  skip normalization factors. The function in the exponential is now given by 
\begin{equation}
G_{1}(x,\lambda)=\int w_{1}(x,\lambda)dx = \sum_{k=1}^{\infty}\lambda^{k}\left( \sum_{\alpha=1}^{k-1}w_{1k\alpha} \frac{x^{\alpha+1}}{\alpha+1} \right).
\end{equation}
As already emphasized above, $u_{1}$ is an edge state for $H_{1}$, but
 not an eigenstate of $H_{0}$.  However,  due to supersymmetry, $\epsilon_{1}(\lambda)$ belongs to the spectrum of $H_{0}$ with eigenstate
\begin{equation}
u_{01}= a^{\dagger}_{0} u^{1}_{n,n-2} \equiv \phi^{0}_{n,n-2}.
\end{equation} 
Using Eq.(\ref{ladderop}) it is easy to check that this state has multiple nodes.

Notice that the recurrence relations for $H_{1}$ in Eqs. (\ref{w1kkm1y}-\ref{e1ky}) are quite similar to those of $H_{0}$ in 
Eqs. (\ref{w0kkm1y}-\ref{e0ky}). The former can be obtained from the latter 
by the replacement $b\to b+1$ everywhere and adding the term 
$2(\alpha+1) w_{0k(\alpha+1)}$ to the relations for $\alpha\le k-2$. These changes reflect the effects of the term $2W_{0}^{\prime}$ in the partner 
potential $v_{1}(x,\lambda)$ in Eq.(\ref{v1h}). 

In  Step 9  of the SEA framework, we construct and solve by the same method the superpartner to 
$H_{1} $, denoted as  $H_{2}$,  whose potential is
\begin{align}
v_{2}(x,\lambda)&=v_{1}(x,\lambda)+2 W^{\prime}_{1}(x,\lambda)=
 \frac{(\ell+2)(\ell+3)}{x^{2}}-\frac{2}{x} + \sum_{k=1}^{\infty}(v_{1k}+2w_{1k}^{\prime}) \lambda^{k}= \sum_{k=0}^{\infty}v_{2k}(x) \lambda^{k}.
 \label{v2h}
\end{align}
For the leading Coulomb term we find
\begin{equation}
w_{20}(x)=\frac{1}{b+2} - \frac{b+2}{x}, \qquad \varepsilon_{20}=-\frac{1}{(b+2)^{2}},  
\end{equation} 
while  for $k\ge 1$ we encounter the following recurrence relations:
\begin{align}
w_{2k(k-1)}&=\frac{b+2}{2}\frac{h_{k}}{k!}   & & \label{w2kkm1y} \\
w_{2k\alpha}&=\frac{b+2}{2}\left[ (2(b+2)+\alpha+1)w_{2k(\alpha+1)} \right. \nonumber \\ 
 &\left. +2(\alpha+1) (w_{0k(\alpha+1)} +w_{1k(\alpha+1)} )-B_{2k\alpha}\right], & \alpha&=k-2, k-1, ...2. \label{w2kay}\\
w_{2k1}&=\frac{b+2}{2}\left[ 2(b+3)w_{2k2}+4 (w_{0k2}+w_{1k2}) \right], & &  \label{w2k1y} \\
\varepsilon_{2k}&=(2b+5)w_{2k1} + 2(w_{0k1} + w_{1k1})+ 2 \delta_{k1}. & & \label{e2ky} 
\end{align}
They yield the solution for the edge eigenstate $u_{2}$ of $H_{2}$ as
\begin{equation}
u_{2}(x,\lambda)= x^{n}e^{-\frac{x}{n}} e^{-G_{2}(x,\lambda)} \equiv \phi^{2}_{n,n-2}(x,\lambda),
\end{equation}
where
\begin{equation}
G_{2}(x,\lambda)=\int w_{2}(x,\lambda)dx = \sum_{k=1}^{\infty}\lambda^{k}\left( \sum_{\alpha=1}^{k-1}w_{2k\alpha} \frac{x^{\alpha+1}}{\alpha+1} \right).
\end{equation}
The eigenvalue $\epsilon_{2}(\lambda)$ is a common eigenstate of $H_{2},H_{1},H_{0}$ and the eigenstates to $H_1$ and $H_{0}$ and  given in their turn as
\begin{align}
u_{12}&=a^{\dagger}_{1}u_{2}=\phi^{1}_{n,n-2}, \\
u_{02}&=a^{\dagger}_{0}a^{\dagger}_{1}u_{2}=\phi^{0}_{n,n-2}.
\end{align}
Finally, in Step 10, the solution to $H_{r}$ is obtained in a similar way. For $k=0$ we find
\begin{equation}
w_{r0}(x)=\frac{1}{b+r} - \frac{b+r}{x}, \qquad \varepsilon_{r0}=-\frac{1}{(b+r)^{2}},  
\label{wr0c}
\end{equation} 
while for $k\ge1$ we encounter the following recurrence relations:
 \begin{align}
w_{rk(k-1)}&=\frac{b+r}{2}\frac{h_{k}}{k!} ,  & & \label{wrkkm1h} \\
w_{rk\alpha}&=\frac{b+r}{2}\left[ (2(b+r)+\alpha+1)w_{rk(\alpha+1)}+2(\alpha+1) \sum_{q=0}^{r-1}w_{qk(\alpha+1)} 
-B_{rk\alpha}\right], & \alpha&=k-2, k-3, ...2, \label{wrkah}\\
w_{rk1}&=(b+r)\left[ (b+r+1)w_{rk2} + 2 \sum_{q=0}^{r-1}w_{qk2}  \right], & &  \label{wrk1h} \\
\varepsilon_{rk}&=(2(b+r)+1)w_{rk1} + 2\sum_{q=0}^{r-1}w_{qk1} + 2 \delta_{k1}. & & \label{erkh} 
\end{align}
They yield the complete solution for the edge eigenstate $u_{r}$ of $H_{r}$,  given by
\begin{align}
u_{r}(x,\lambda)&= x^{n}e^{-\frac{x}{n}} e^{-G_{r}(x,\lambda)}\equiv \phi^{r}_{n,n-1-r}(x,\lambda), 
\end{align}
where
\begin{equation}
G_{r}(x,\lambda)=\int w_{r}(x,\lambda)dx = \sum_{k=1}^{\infty}\lambda^{k}\left( \sum_{\alpha=1}^{k-1}w_{rk\alpha} \frac{x^{\alpha+1}}{\alpha+1} \right).
\end{equation}
This is not an eigenstate to $H_{0}$ but supersymmetry connects the chain of Hamiltonians $\{ H_{0},H_{1},....,H_{r} \}$, thus making
$\epsilon_{r}(\lambda)$ a common eigenvalue. The corresponding eigenstate of
$H_0$ is given by
\begin{equation}
u_{0r}= a^{\dagger}_{0}a^{\dagger}_{1}...a^{\dagger}_{r-2}a^{\dagger}_{r-1} u_{r}\equiv \phi^{0}_{n,n-1-r}.
\end{equation} 
The procedure  terminates for $r=n-1$ with  reaching and solving for the edge eigenstate state $u_{n-1}\equiv\phi^{n-1}_{n0}$ to $H_{n-1}$,  which then  yields 
the solution $\phi^{0}_{n0}$ of the Hulth\'en potential $H_{0}$ as
\begin{equation}
u_{0(n-1)}= a^{\dagger}_{0}a^{\dagger}_{1}...a^{\dagger}_{n-3}a^{\dagger}_{n-2} u_{n-1}\equiv \phi^{0}_{n,0}.
\end{equation} 
In our calculation we skipped normalization factors for the sake of simplicity,
so at the end  every state needs to be properly normalized.

Summarizing this long calculation, the eigenstates and eigenvalues for the Hulth\'en potential $H_{0}$ are given by
\begin{align}
\phi^{0}_{n,\ell}&=N_{n,\ell}(\lambda) a^{\dagger}_{0}a^{\dagger}_{1}...a^{\dagger}_{r-2}a^{\dagger}_{r-1} u_{r}, \label{phinlh}\\
\epsilon_{n\ell}(\lambda)&=\sum_{\lambda=0}^{\infty} \left[\left( 2n+1\right)w_{rk1} + 2\sum_{q=0}^{r-1}w_{qk1}   +2\delta_{k1}\right]\lambda^{k},  \label{enlh}
\end{align} 
where $\ell=n-1-r$.

Our procedure is graphically illustrated  in Fig. \ref{seapich} for the solution of the level $n=5$, where we use the notation 
$\phi^{r}_{n\ell}=|n\ell\rangle_{r}$. We first solve the Riccati equation for $H_{0}$ using the expansion
of the potential $v_{0}(x,\lambda)$, to find the energy $\epsilon_{0}(\lambda)$ and superpotential $W_{0}(x,\lambda)$.
The obtained solution corresponds to the $|5,4\rangle_{0}$ edge state. 
Then we use supersymmetry to  build and solve likewise the superpartner $H_{1}$, whose edge state turns out to be $|5,3\rangle_{1}$. 
The eigenstate $|5,3\rangle_{0}$ of $H_{0}$ of our interest, is obtained by the action of the
operator $a^{\dagger}_{0}$ on $|5,3\rangle_{1}$. The remaining $|5,\ell\rangle_{0}$ states for $\ell < 3$ are obtained constructing likewise  
the super-partner Hamiltonians $\{H_{2}, H_{3},H_{4}\}$, solving them by the same procedure,  which yields the respective 
$\{ |5,2\rangle_{2},|5,1\rangle_{3},|5,0\rangle_{4}\}$ edge states. The eigenstates 
$\{ |5,2\rangle_{0},|5,1\rangle_{0},|5,0\rangle_{0}\}$ of $H_{0}$ 
are obtained  by the successive action of the $a^{\dagger}_{3}, a^{\dagger}_{2},  a^{\dagger}_{1}, a^{\dagger}_{0}$ factorization 
operators on the former states. In this manner, we solve completely for all the eigenstates of $H_{0}$ in the  $n=5$ level. 
The same procedure is applied to every $n$ level to achieve the complete analytical solution for the Hulth\'en potential. 

It is important to emphasize that, unlike the Coulomb case, for the Hulth\'en potential the supersymmetric  Hamiltonians $\{ H_{0}, H_{1}, ... , H_{r} \}$
do not constitute a family of shape invariant potentials, i.e., they can not be related to each other  just by a  parameter change. 
A similar observation has been reported regarding the Yukawa potential in 
\cite{Napsuciale:2020ehf,Napsuciale:2021qtw}. There,  the property 
of shape invariance was lost already  beyond the order of $\lambda^{2}$ in the 
potential's expansion. Shape invariance is a powerful property  in the Liouvillian solvable SUSYQM,
but it is not  required  by the No Liouvillian solvable one, which is our case.
Stated differently, our formalism goes beyond shape invariance and thus makes use of the full power of supersymmetry.
It is also worth mentioning that as a byproduct of the aforementioned calculation the complete analytical solutions of all the members of the chain of 
supersymmetric  Hamiltonians $\{ H_{0}, H_{1}, ... , H_{r} \}$ were obtained, not only those of the genuine Hulth\'en Hamiltonian $H_{0}$. Furthermore, 
the calculation at a given order $\lambda^{k}$ yields the complete exact solution of the expansion of any one of the Hamiltonians in the chain to this order. 
Therefore, we actually could obtain  systematically the analytical solutions to an infinite set of Hamiltonians 
$\{ H^{(k)}_{0}, H^{(k)}_{1}, ... , H^{(k)}_{r} \}$, $k=0,1,...\infty$. Similar considerations apply to the supersymmetric Hamiltonians to
any arbitrary potential solved by the SEA approach. 

\begin{figure}
\begin{center}
\begin{tikzpicture}
\draw[gray, thick,->] (0, 0)--(0,5) node[above]{$n$} ;
\draw[gray, thick,->] (0, 0)--(8,0)  node[right]{$l$} ;
\draw[gray, thick,->] (0, 0)--(-3,-3) node[below]{$r$}  ;

\draw[scalar] (-4,0)--(-3,1)  ;
\draw[scalar] (-3,1)--(-2,2)  ;
\draw[scalar] (-2,2)--(-1,3)  ;
\draw[scalar] (-1,3)--(0,4)  ;

\draw[fermion] (8.7,4.6)-- (6.3,3.7);
\draw(8,4)node {$SUSY$} ;
\draw[fermion] (5.7,3.4)-- (3.3,2.4) ;
\draw(5,2.7)node {$SUSY$} ;
\draw[fermion] (2.7,2.2)-- (0.2,1.2) ;
\draw(1.8,1.5)node {$SUSY$} ;
\draw[fermion] (-0.4,1)-- (-2.6,0.1) ;
\draw(-1,0.3)node {$SUSY$} ;

\draw[black,ultra thick] (7.8,4.8)--(8.2,4.8)  node[above ]{{\color{blue}$|5,4\rangle_{0}$}} ; 
\draw(9,4.8)node{$H_{0}$};
\draw[black,ultra thick] (5.8,4.6)--(6.2,4.6)  node[above ]{$|5,3\rangle_{0}$} ; 
\draw[black,ultra thick] (3.8,4.4)--(4.2,4.4)  node[above ]{$|5,2\rangle_{0}$} ; 
\draw[black,ultra thick] (1.8,4.2)--(2.2,4.2)  node[above ]{$|5,1\rangle_{0}$} ; 
\draw[black,ultra thick] (-0.2,4)--(0.2,4)  node[above ]{$|5,0\rangle_{0}$} ; 


\draw[black,ultra thick] (4.8,3.6)--(5.2,3.6)  node[above ]{{\color{blue}$|5,3\rangle_{1}$}} ; 
\draw(6,3.6)node{$H_{1}$};
\draw[black,ultra thick] (2.8,3.4)--(3.2,3.4)  node[above ]{$|5,2\rangle_{1}$} ; 
\draw[black,ultra thick] (0.8,3.2)--(1.2,3.2)  node[above ]{$|5,1\rangle_{1}$} ; 
\draw[black,ultra thick] (-1.2,3)--(-0.8,3)  node[above ]{$|5,0\rangle_{1}$} ; 


\draw[black,ultra thick] (1.8,2.4)--(2.2,2.4)  node[above ]{{\color{blue}$|5,2\rangle_{2}$}} ; 
\draw(3,2.4)node{$H_{2}$};
\draw[black,ultra thick] (-0.2,2.2)--(0.2,2.2)  node[above ]{$|5,1\rangle_{2}$} ; 
\draw[black,ultra thick] (-2.2,2)--(-1.8,2)  node[above ]{$|5,0\rangle_{2}$} ;


\draw[black,ultra thick] (-1.2,1.2)--(-0.8,1.2)  node[above ]{{\color{blue}$|5,1\rangle_{3}$}} ; 
\draw(-0.2,1.2)node{$H_{3}$};
\draw[black,ultra thick] (-3.2,1)--(-2.8,1)  node[above ]{$|5,0\rangle_{3}$} ; 

\draw[black,ultra thick] (-4.2,0)--(-3.8,0)  node[above ]{{\color{blue}$|5,0\rangle_{4}$}} ; 
\draw(-3,0)node{$H_{4}$};

\draw[scalar] (-1,1.2)--(0,2.2)  ;
\draw[scalar] (0,2.2)--(1,3.2)  ;
\draw[scalar] (1,3.2)--(2,4.2)  ;

\draw(-0.6,1.95)node{$a^{\dagger}_{2}$};
\draw(-2.6,1.75)node{$a^{\dagger}_{2}$};
\draw(-3.6,0.75)node{$a^{\dagger}_{3}$};

\draw[scalar] (2,2.4)--(3,3.4)  ;
\draw[scalar] (3,3.4)--(4,4.4)  ;
\draw(-1.6,2.75)node{$a^{\dagger}_{1}$};
\draw(0.4,2.95)node{$a^{\dagger}_{1}$};
\draw(2.4,3.15)node{$a^{\dagger}_{1}$};

\draw[scalar] (5,3.6)--(6,4.6)  ;
\draw(-0.6,3.75)node{$a^{\dagger}_{0}$};
\draw(1.4,3.95)node{$a^{\dagger}_{0}$};
\draw(3.4,4.15)node{$a^{\dagger}_{0}$};
\draw(5.4,4.35)node{$a^{\dagger}_{0}$};

\end{tikzpicture}
\end{center}
\caption{Connections between states and operators for the superpartners $\{ H_{0},H_{1},H_{2},H_{3},H_{4}\}$ 
in the calculation of the $|n,\ell\rangle_{0}$ eigenstates with $n=5$ of the Coulomb Hamiltonian $H_{0}$ using  
supersymmetry and the edge states (see text) $|n,\ell\rangle_{r}$ with $\ell=n-1-r$ for each superpartner $H_{r}$.}
\label{seapich}
\end{figure}

The results for the energy levels $\epsilon_{n\ell}(\lambda)$ in Eq.(\ref{enlh}) can be written in terms of $n$ and $\ell$. It turns out 
that they depend only on $n^{2}$ and $L^{2}=\ell(\ell+1)$ according to
\begin{equation}
\epsilon_{n\ell}(\lambda)=\sum_{k=0}^{\infty}\varepsilon_{k} (n^{2}, L^{2})\lambda^{k}.
\end{equation}
The coefficients $\varepsilon_{k} (n^{2}, L^{2})$ have long expressions for large $k$ thus we list them here only up to $k=10$ 
\begin{align}
 \varepsilon_{0}(n^{2}, L^{2})&=-\frac{1}{n^{2}}, \\
 \varepsilon_{1}(n^{2}, L^{2})&=1 ,\\
\varepsilon_{2}(n^{2}, L^{2})&=-\frac{1}{12}(3n^{2}-L^{2}), \\
\varepsilon_{3}(n^{2}, L^{2})&= 0,\\
\varepsilon_{4}(n^{2}, L^{2})&= -\frac{n^{2}L^{2}}{480}(5n^{2} -3 L^{2}+1), \\
 \varepsilon_{5}(n^{2}, L^{2})&=0.\\
 \varepsilon_{6}(n^{2}, L^{2})&= -\frac{n^{4}L^{2}}{483840}(315 n^{4}  + 406 n^{2}L^{2} - 465 L^{4}+ 273 n^{2} +290 L^{2} -120), \\
 \varepsilon_{7}(n^{2}, L^{2})&=0.\\
 \varepsilon_{8}(n^{2}, L^{2})&=  -\frac{n^{6}L^{2}}{116121600} (9282 n^{6}  + 15015 n^{4}L^{2} +  9000 n^{2} L^{4} - 21553 L^{6} + 18480 n^{4}  \nonumber \\
 &+23445 n^{2}L^{2} +19292 L^{4} -  23082 n^{2} -14952 L^{2} +7560) ,\\
 \varepsilon_{9}(n^{2}, L^{2})&=0.\\
 \varepsilon_{10}(n^{2}, L^{2})&= -\frac{n^{8}L^{2}}{20437401600} (250767 n^{8} +582120 n^{6}L^{2}+370370n^{4}L^{4}+101816 L^{6}n^{2} - 821745 L^{8}  \nonumber \\ 
 &+ 898590 n^{6}+1981980n^{4}L^{2} +899646 n^{2}L^{4} +952860 L^{6}  ) \nonumber \\ 
 &- 2219217 n^{4} -1889316n^{2}L^{2} -1015500L^{4} +1956900 n^{2} +1030320 L^{2}-604800).
\end{align}
Notice that all the coefficients for odd $k\ge3$ vanish and for $k\ge 4$ the even coefficients are proportional to $\ell(\ell+1)$ thus, for the $\ell=0$ states 
the contributions beyond  $k= 2$ are nule and the energy levels gain the exact form
\begin{equation}
\epsilon_{n,0}(\lambda)=-\frac{1}{n^{2}} +\lambda -\frac{n^{2}}{4}\lambda^{2}=-\left(\frac{1}{n} - \frac{n\lambda}{2}\right)^{2}.
\end{equation}
This result has been previously obtained in \cite{Lam:1971,Flugge:1999} where a closed solution for the $\ell=0$ states is given. 
It is interesting to explicitly obtain the closed solutions in this case, since our power series solution must correspond to these closed form solutions in this
particular case. Following  \cite{Lam:1971} we write
\begin{equation}
\phi(x,\lambda)=e^{-\lambda b x}\eta(x,\lambda),
\end{equation}
where $b$ is a free parameter. The Schr\"{o}dinger equation for $\ell=0$ states yields the following equation for $\eta$
\begin{equation}
\left[ \left(-\lambda b+\frac{d}{dx}\right)^{2}+\frac{2\lambda e^{-\lambda x}}{1-e^{-\lambda x}} +\epsilon \right] \eta(x,\lambda)=0.
\end{equation}
Changing variable to $y=1-e^{-\lambda x}$ gives
\begin{equation}
(1-y)^{2}\eta^{\prime\prime}-(2b+1)(1-y)\eta^{\prime}+\left[\frac{2(1-y)}{\lambda y} +\frac{\epsilon}{\lambda^{2}}+ b^{2}\right]\eta=0.
\end{equation}
This equation simplifies with the choice
\begin{equation}
 b^{2}=-\frac{\epsilon}{\lambda^{2}},
\end{equation}
which yields
\begin{equation}
y(1-y)\eta^{\prime\prime}-(2b+1)y\eta^{\prime} + \frac{2}{\lambda } \eta=0.
\label{phieq}
\end{equation}
Expanding in powers of $y$ 
\begin{equation}
\eta(y,\lambda)=\sum_{k=0}^{\infty}c_{k}(\lambda)y^{k},
\end{equation}
and inserting this series in Eq. (\ref{phieq}) gives  $ c_{0}(\lambda)=0$ together with  the following recurrence relations for $k\ge1$:
\begin{equation}
c_{k+1}(\lambda)=\frac{k(k-1) + (2b+1)k-\frac{2}{\lambda}}{k(k+1)} c_{k}(\lambda).
\end{equation}
Obviously, the series  diverges for $x\to\infty$, and one can find  solutions only if  $c_{k+1}$ nullifies for some $k=n$, in which case
a quantization condition is obtained as
\begin{equation}
b_{n}=\frac{1}{n\lambda}-\frac{n}{2}.
\end{equation}
Thus, the energy levels for the $\ell=0$ states of the Hulth\'en potential are 
obtained as
\begin{equation}
\epsilon_{n0}(\lambda)=-(\lambda b_{n})^{2}=-\left(\frac{1}{n} - \frac{n\lambda}{2}\right)^{2}.
\end{equation}
The explicit form of the states in terms of $y$ are then found as
\begin{equation}
\phi_{n0}(y)=(1-y)^{\frac{1}{n\lambda}-\frac{n}2} \eta_{n}(y,\lambda), 
\end{equation}
where $\eta_{n}(y,\lambda) $ are polynomials in $y$ of degree $n$. These solutions are normalizable only for $0\leq\lambda < 2/n^{2}$. 
The polynomials for the normalized ground and first excited states read:
\begin{align}
\eta_{1}(y,\lambda)&= \frac{\sqrt{4-\lambda^{2}}}{\lambda}y,  \\
\eta_{2}(y,\lambda)&= \frac{\sqrt{1-4\lambda^{2}}}{\sqrt{2}\lambda}\left(y- \frac{1}{2}\left(\frac{1}{\lambda}+1\right)y^{2}\right).  
\end{align}
In terms of the variable $x$ we find
\begin{align}
\phi_{10}(x,\lambda)&=2\frac{\sqrt{4-\lambda^{2}}}{\lambda}\, e^{-x} \sinh\frac{\lambda x}{2}, \\
\phi_{20}(x,\lambda)&=\sqrt{2}\frac{\sqrt{1-4\lambda^{2}}}{\lambda}e^{-\frac{1-\lambda}{2}x}\sinh\frac{\lambda x}{2}
\left(1-(\frac{1}{\lambda}+1)e^{-\frac{\lambda x}{2}}\sinh\frac{\lambda x}{2}\right).
\end{align}
Expanding the states in powers of $\lambda$,  the series obtained within 
 the SEA framework are recovered. This can be straightforwardly 
verified for the ground state 
at the level of the logarithmic form of Schr\"odinger's equation for $\ell=0$, 
\begin{equation}
W^{2}_{0}(x,\lambda) - W^{\prime}_{0}(x,\lambda)=v_{0}-\epsilon_{0}=\lambda(1-\coth\frac{\lambda x}{2}) -\epsilon_{0},
\end{equation}
solved by
\begin{equation}
W_{0}(x,\lambda)=1-\frac{\lambda}{2}\coth\frac{\lambda x}{2}.
\end{equation}
The $ W_{0}(x,\lambda)$ expansion then  coincides with our solution  
in Eq(\ref{W0h}) for the case $n=1$. The  complete solution 
to the Hulth\'en potential obtained here includes the results in \cite{Lam:1971,Flugge:1999} for the $\ell=0$ states and also the 
${\cal{O}}(\lambda^{6})$ results obtained in \cite{Lai:1980} for arbitrary $\ell$.
  
\begin{figure}%
\centering
\includegraphics[width=7cm]{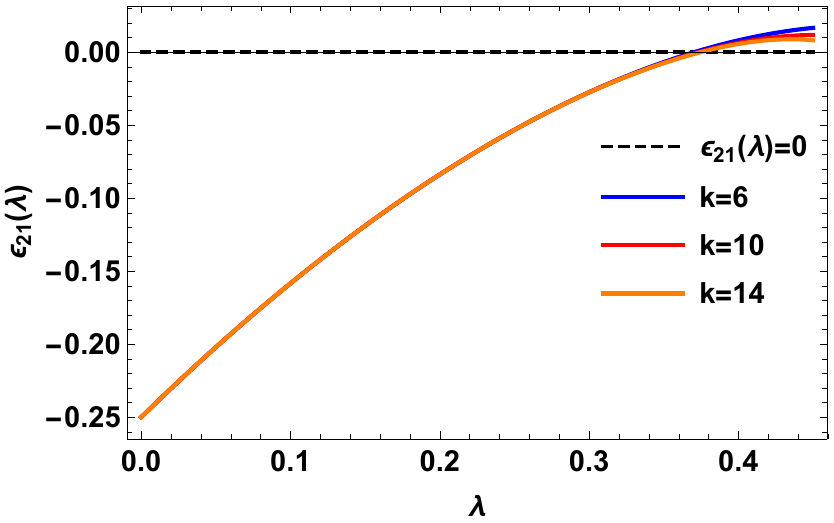} %
\qquad
\includegraphics[width=7cm]{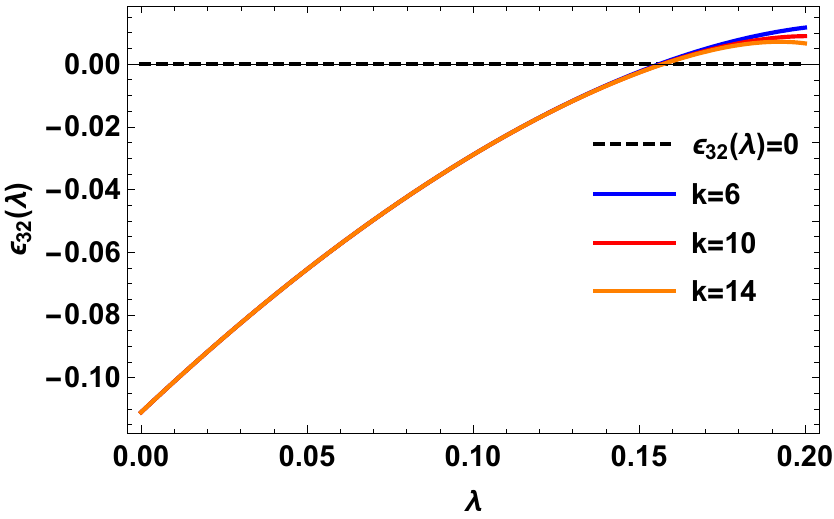} %
\qquad
\includegraphics[width=7cm]{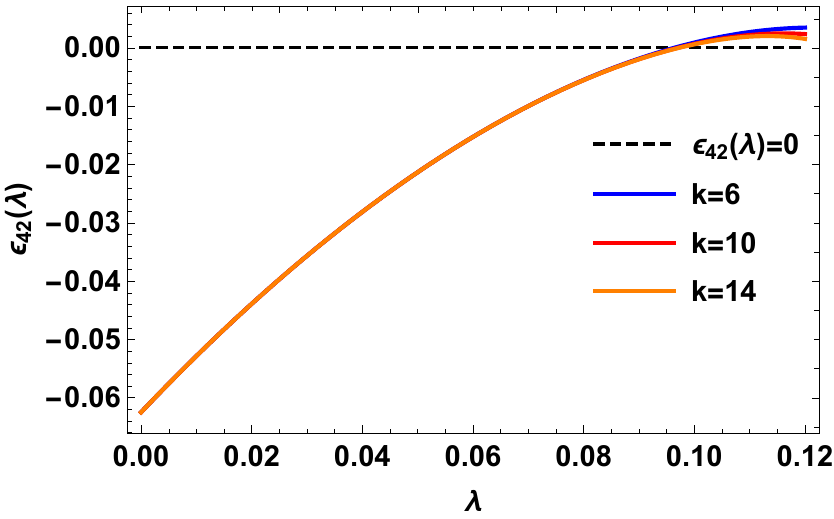} %
\qquad
\includegraphics[width=7cm]{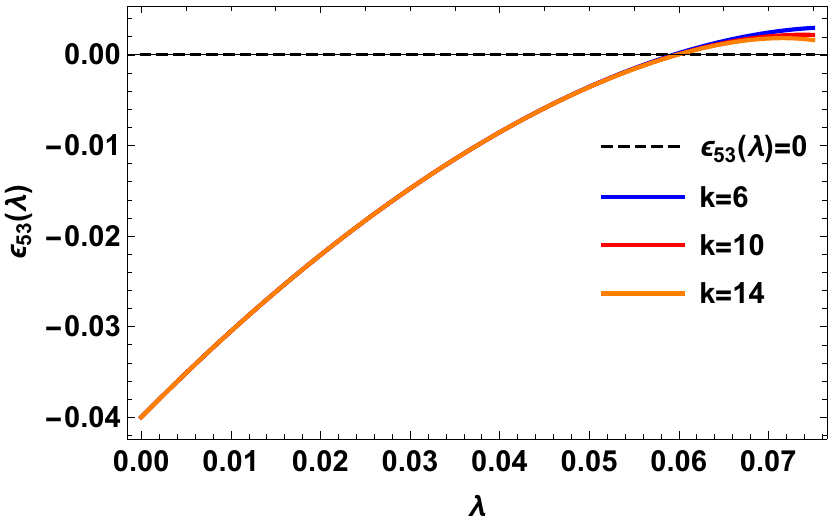} %
\caption{Lowest energy levels $\epsilon_{n\ell}(\lambda)$ with $\ell\neq 0$ calculated up to $\lambda^{k}$ for $k=6,10,14$.}%
\label{Enlk}%
\end{figure}

In Fig. \ref{Enlk} we plot $\epsilon_{n\ell}(\lambda)$ for $n\ell=21,32,42$ and $53$, considering the expansions 
up to $\lambda^{k}$ with $k=6,10,14$. We can see that the series are convergent in the whole range of values of $\lambda$ where bound 
states exist, i.e., below the critical value $\lambda_{c}$ where $\epsilon_{n\ell}(\lambda_{c})=0$. Similar results are obtained for all the 
$\epsilon_{n\ell}$ levels. 

Using the expansion up to $k=14$ we estimate the critical values with an 
accuracy at the percent level. More precise results can be obtained if we reconstruct the function $\epsilon_{n\ell}(\lambda)$ using methods 
like the Pad\'{e} approximants. In Table \ref{lambdac} we list the precise results for the critical screening lengths obtained from the $[15/14]$ 
and $[14/14]$ Pad\'{e} approximants (which requires to calculate the Taylor series of $\epsilon_{n\ell}(\lambda)$ up to $\lambda^{30}$). The 
uncertainty in the calculation is taken from the difference in the result obtained using the $[15/14]$ and $[14/14]$ Pad\'{e} approximants.  

\begin{figure}%
\centering
\includegraphics[width=7cm]{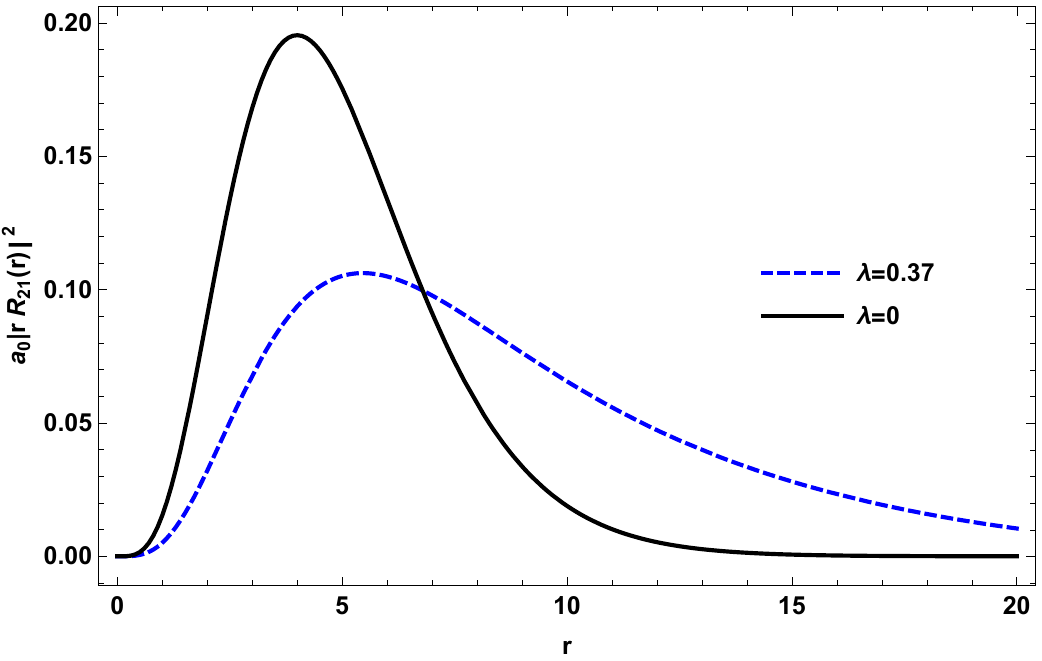} %
\qquad
\includegraphics[width=7cm]{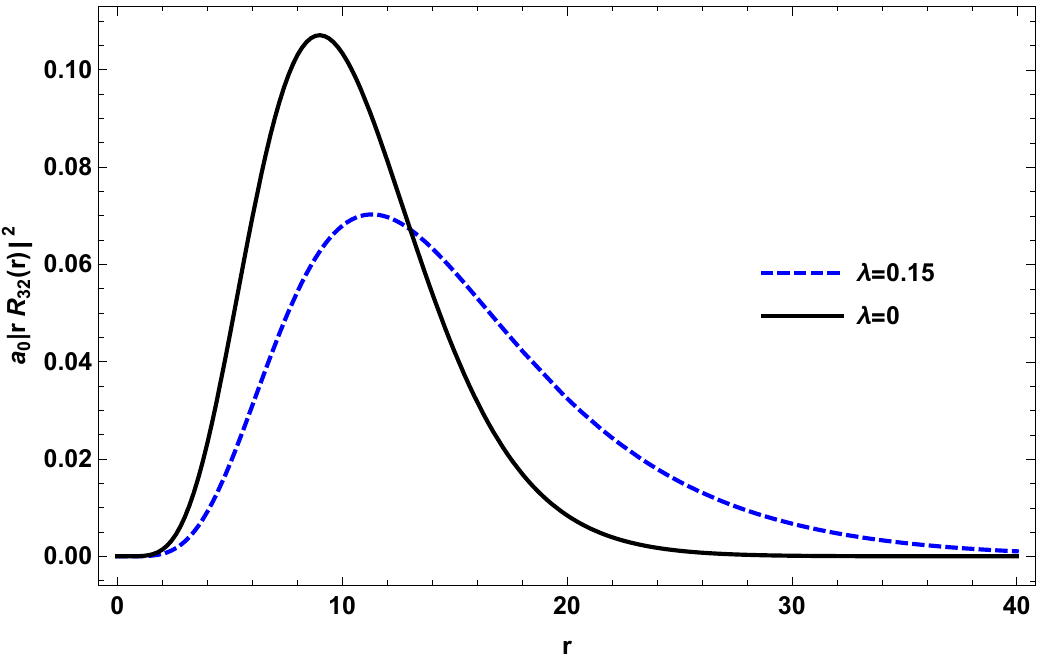} %

\includegraphics[width=7cm]{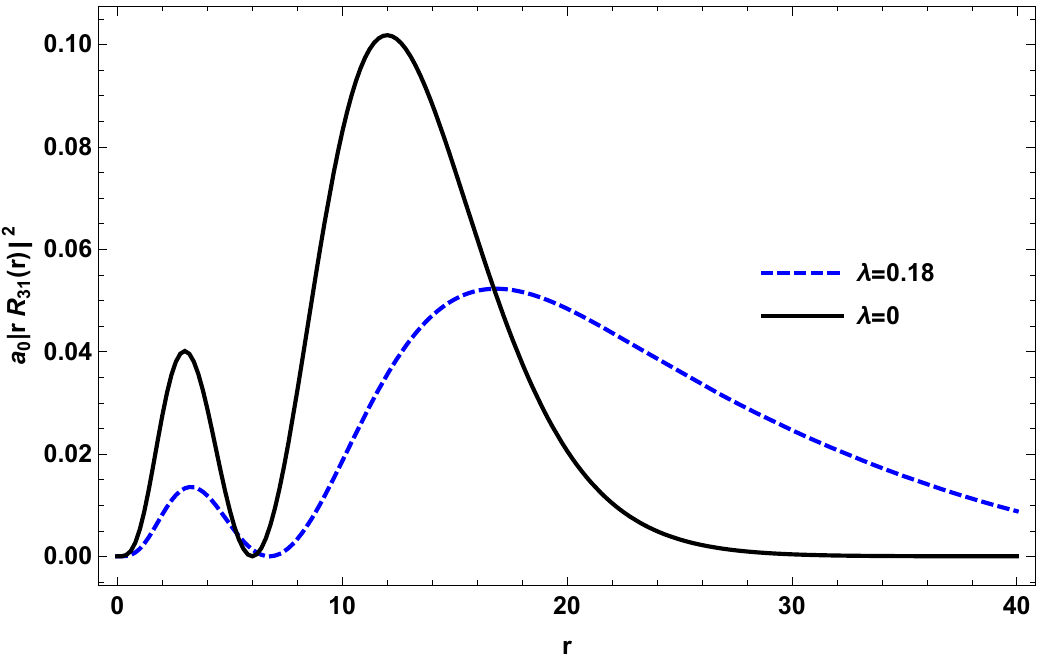} %
\qquad
\includegraphics[width=7cm]{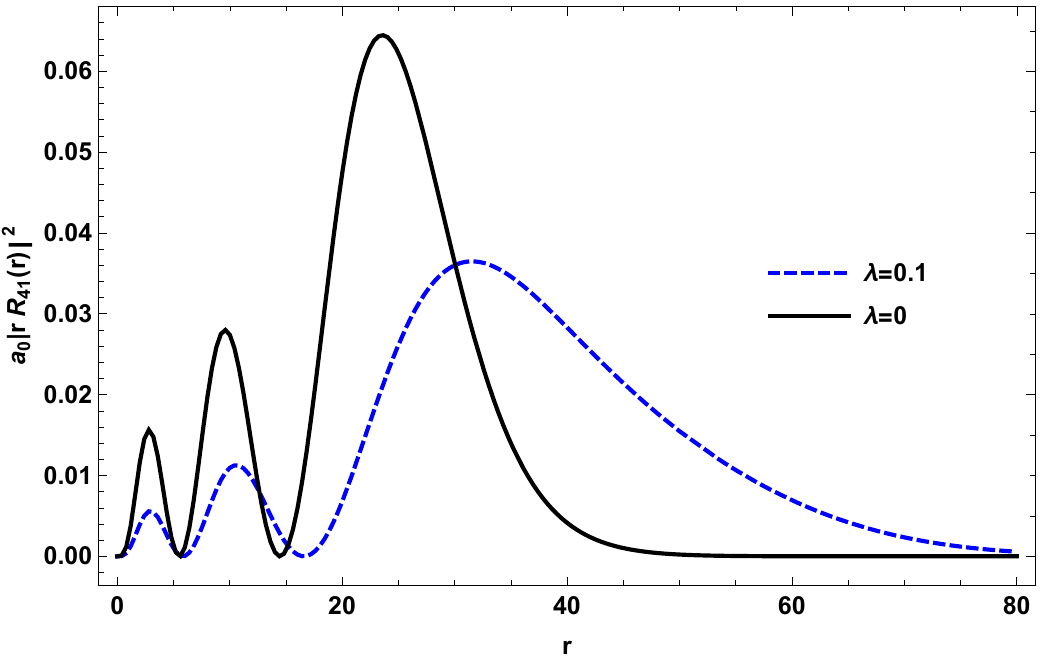} %
\caption{Radial probabilities for the $\ell\neq 0$ lowest lying states of the Hulth\'en potential for $\lambda=0$ (Coulomb case) and a value of 
        $\lambda$ close the the critical value for each state.}%
\label{ProbH}%
\end{figure}

As to the eigenstates, in Fig. \ref{ProbH} we plot the radial probabilities for the lowest lying $\ell\neq 0$ calculated with the $[5/5](\lambda)$ 
Pad\'e approximant for values of $\lambda$ close to the critical values, together with the Coulomb probabilities. We can see in these plots 
the beginning of the delocalization of the states for $\lambda$ close to the critical values, even though the peaks in the radial probability are 
only slightly shifted towards larger radius, the shift being more pronounced for excited states with low values of $\ell$.

\begin{table}[!ht]
\centering
 {
    \centering
    \begin{tabular}{|c|c|c|}
\hline
   $n$ & $\ell$ & $\lambda_{c}$\\
\hline
   1&0&2\\ 
\hline
   2&0&1/2\\ 
   2&1& 0.3767388(1)\\
\hline
   3&0&2/9\\ 
   3&1&0.18638519(1)\\ 
   3&2&0.1576540(1)\\ 
\hline
   4&0&1/8\\ 
   4&1&0.11042423(5)\\ 
   4&2&0.09755514(4)\\ 
    4&3&0.08640416(2)\\ 
\hline
   5&0&2/25\\ 
   5&1&0.07281399(4) \\ 
   5&2&0.06609952(3)\\ 
   5&3&0.05997137(1)\\ 
   5&4&0.054505130(5)\\ 
 \hline
   \end{tabular}
  }
{
    \centering
    \begin{tabular}{|c|c|c|}
\hline
   $n$ & $l$ & $\lambda_{c}$\\
\hline
       6&0&1/18\\ 
   6&1&0.05154187(2)\\ 
   6&2&0.04765376(2)\\ 
   6&3&0.04397303(1)\\ 
   6&4&0.040584332(5)\\
   6&5&0.037504108(2)\\ 
\hline
   7&0&2/49\\ 
   7&1&0.03836901(2)\\ 
   7&2&0.03594088(1)\\ 
   7&3&0.033579387(7)\\ 
   7&4&0.031352334(4)\\
   7&5&0.029284146(2)\\ 
   7&6&0.027378996(1)\\ 
\hline
    \end{tabular}
  }
{
    \centering
     \begin{tabular}{|c|c|c|}
\hline
   $n$ & $l$ & $\lambda_{c}$\\
\hline
8&0&1/32\\ 
   8&1&0.02965680(2)\\ 
   8&2&0.02805166(1) \\ 
   8&3&0.02645746(1)\\ 
   8&4&0.024925430(3)\\
   8&5&0.023478153(2)\\ 
   8&6&0.022124095(1)\\ 
   8&7&0.0208642596(4)\\ 
\hline
   9&0&2/81\\ 
   9&1&0.02360076(1) \\ 
   9&2&0.02249094(1)\\ 
   9&3&0.021370275(5)\\ 
   9&4&0.020276903(3)\\
   9&5&0.019229555(2)\\ 
   9&6&0.018237044(1)\\ 
   9&7&0.0173026475(5)\\ 
   9&8&0.0164264743(2)\\ 
   \hline
    \end{tabular}
  }
\caption{Critical screening lengths for the Hulth\'en potential, calculated using the reconstruction of $\epsilon_{n\ell}(\lambda)$ 
with the $[15/14]$ and $[14/14]$ Pad\'{e} approximants.}
\label{lambdac}
\end{table}

We will release with the published version of this paper a freely available code for the solution of the recurrence relations which provides the solutions 
for the energy levels and normalized eigenstates to the desired order $\lambda^{k}$, along with some details of the phenomenology for 
the Hulth\'en potential.

\section{One dimensional anharmonic oscillator  potential within the SEA approach and perturbation theory}

\subsection{Anharmonic oscillator potential}
Every physical systems behave like  harmonic oscillators for small motion around given
equilibrium points, characterized by  local minima in the 
corresponding potentials. For parity invariant interactions, the leading correction to this simple behavior is the anharmonic potential which, 
for a one-dimensional system, has the form of $\lambda x^{4}$. The calculation of the corrections to the energy levels is of a  primary concern for 
several branches of natural sciences and in the past there have been many attempts to solve this problem. The analytic structure of the 
{\emph{ground state}} energy using the WKB approximation and the calculation of the Taylor series for the Rayleig-Schr\"{o}dinger anharmonic 
corrections for this level have been reported  by
 \cite{Bender:1969si} up to the very high order ($\lambda^{75}$) in the perturbation theory, definitively knotting  though on diverging series.
A rigorous derivation of the analytic structure of the ground state energy and the asymptotic behavior at $\lambda\to \infty$  were given in 
\cite{Simon:1970mc}, where the asymptotic nature of the series was established. Precise results for this level were given also in 
\cite{Loeffel:1969rdm}, where a reconstruction of the ground state energy was performed  employing the $[20/20]$ Pad\'{e} approximant, 
with the coefficients of the Taylor series found  in \cite{Bender:1969si}.  For other techniques for obtaining approximate solutions to the 
anharmonic oscillator for the ground and excited states see \cite{Hioe:1978jj,Amore:2004} and references there in.
 Upper and lower bounds on the actual values of the first energy 
levels have been set in  \cite{Bazley:1961}. Moreover, the ground state energy  for the $D$-dimensional harmonic oscillator and the corresponding 
anharmonic corrections up to the order of $\lambda^{70}$ 
to it as following from the  LPT framework, have been  worked out in \cite{Dolgov:1979hv}. 
In summary, though one can find  approximate calculations of the anharmonic corrections to the harmonic oscillator,  a systematic complete 
solution still seems to be an open problem. In the following we will make use of  the supersymmetric expansion algorithm to provide such a solution. 

The one-dimensional anharmonic potential reads
\begin{equation}
V(z,\lambda_{0})=\frac{1}{2}m\omega^{2}z^{2} + \frac{\lambda_{0}}{4}z^{4}.
\end{equation}
Using the typical scale of the leading harmonic oscillator term, $a=\sqrt{\hbar/m\omega}$ we obtain
\begin{equation}
v_{0}(x)=x^{2}+ \lambda x^{4}, \qquad \epsilon_{0}(\lambda)=\frac{2E}{\hbar \omega},
\end{equation}
where $\lambda=\lambda_{0}\hbar/m^{2}\omega^{3}$.
We can view this potential as a power expansion in $\lambda$ with the coefficients 
\begin{align}
v_{0k}(x) =\left \{
\begin{array}{r}
x^{2} \quad \textit{for} \quad k=0,\\ 
x^{4}\quad \textit{for} \quad k= 1 \\
0 \quad \textit{for} \quad k\ge 2 .
  \end{array}
  \right.
\label{v0kao}
\end{align}
The cascade of equations for $w_{k}$ in this case emerges as
\begin{align}
k&=0:  & w_{00}^{2} - w'_{00} &= x^{2} - \varepsilon_{00}, \label{w00ao} \\
k&=1:  & 2 w_{00} w_{01}- w'_{01} &=x^{4} -\varepsilon_{01}, \label{w01ao} \\
k&\ge 2:  & 2 w_{00} w_{0k}- w'_{0k} &= -B_{0k} -\varepsilon_{0k}.  \label{w0kao} 
\end{align}
These equations have polynomial solutions which can be easily worked out. 
In particular, the solutions up to $k=5$ are 
\begin{align}
w_{00}(x)&= x, &  \varepsilon_{00}&=1, \label{solao00}\\
w_{01}(x)&= \frac{3}{4}x + \frac{1}{2}x^{3},  & \varepsilon_{01}&=\frac{3}{4} ,\\
w_{02}(x)&= -\frac{21}{16}x - \frac{11}{16}x^{3}- \frac{1}{8}x^{5}, & \varepsilon_{02}&=-\frac{21}{16}, \\
w_{03}(x)&= \frac{333}{64}x + \frac{45}{16}x^{3} + \frac{21}{32}x^{5} + \frac{1}{16}x^{7}, & \varepsilon_{03}&= \frac{333}{64},\\
w_{04}(x)&= -\frac{30885}{61024}x - \frac{8669}{512}x^{3} - \frac{1159}{256}x^{5} - \frac{163}{256}x^{7}- \frac{5}{128}x^{9},  
& \varepsilon_{04}&= -\frac{30885}{1024}, \\
w_{05}(x)&= \frac{916731}{4096}x + \frac{33171}{256}x^{3} + \frac{19359}{512}x^{5} + \frac{823}{128}x^{7}  + \frac{319}{512}x^{9} 
+ \frac{7}{256}x^{11} , 
& \varepsilon_{05}&= \frac{916731}{4096}. \label{solao05}
\end{align}
These results suggest that the coefficients in the expansion of the the super-potential for the first solution to $H_{0}$ are of the form
\begin{equation}
w_{0k}(x)=\sum_{\alpha=1}^{2k+1}w_{0k\alpha}x^{\alpha}.
\label{w0kaogen}
\end{equation}
Inserting Eq.(\ref{w0kaogen}) in Eq. (\ref{w0kao}), we find that for $k\geq 2$, the coefficients $w_{0k\alpha}$ vanish for even $\alpha$,
and for odd $\alpha$ they satisfy the following recurrence relations 
\begin{align}
w_{0 k (2k+1)}  & =-\frac{1}{2}B_{0k(2k+2)},
\end{align}
while for $\alpha=2k,2k-2,...2$ and for the energy coefficients we obtain
\begin{align}
w_{0 k (\alpha-1)}  & = \frac{1}{2}\left(  \left(  \alpha+1\right)  w_{0  k(\alpha+1)} -B_{0k\alpha}\right) ,  \\
\varepsilon_{0k}  & =w_{0k1}.
\end{align}

Skipping normalization factors, the first solution to $H_{0}$ will be
\begin{equation}
u_{0}(x,\lambda)= e^{-G_{0}(x,\lambda) } , \qquad \epsilon_{0}(\lambda)=\sum_{k=0}^{\infty}\varepsilon_{0k}\lambda^{k},
\end{equation}
where
\begin{equation}
G_{0}(x,\lambda)=\int w_{0}(x,\lambda)dx = \sum_{k=1}^{\infty}\lambda^{k}\left( \sum_{\alpha=1}^{2k+1}w_{0k\alpha} \frac{x^{\alpha+1}}{\alpha+1} \right).
\end{equation}
Similarly to the Hulth\'en's potential case, we get more information from the first few terms in the expansion in $\lambda$ for this solution. Indeed, the leading 
term ($k=0$) corresponds to the ground state of the harmonic oscillator, thus this solution is the ground state of $H_{0}$.
We conclude that, unlike the case of the screening potentials, the anharmonic Hamiltonian $H_{0}$ has only one nodeless state and it 
is the ground state $u_{0}(x,\lambda)$. 

The calculation of the excited states requires to construct the set of supersymmetric partners $\{ H_{1},H_{2},...,H_{r}\}$. We skip the details 
of the by now familiar calculations and present only the results for the recurrence relations of $H_{r}$. 
The coefficients in the expansion of the the super-potential for $H_{r}$ are of the form
\begin{equation}
w_{rk}(x)=\sum_{\alpha=1}^{2k+1}w_{rk\alpha}x^{\alpha}.
\end{equation}
For $k=0,1$ the solutions are given by
\begin{align}
w_{r 0}  & =x   , & \varepsilon_{r0}&= 2r+1, \label{wr0ao}\\
w_{r 1}  & = \frac{3}{4}(2r+1) x +\frac{1}{2}x^{3}  , & \varepsilon_{r1}&= \frac{3}{2}(r^{2}+r+\frac{1}{2}).
\end{align}
For $k\geq 2$, the coefficients $w_{rk\alpha}$ also vanish for even $\alpha$ and for odd $\alpha$ they satisfy the following recurrence relations 
\begin{align}
w_{r k (2k+1)}  & =-\frac{1}{2} B_{rk(2k+2)},
\end{align}
while for $\alpha=2k,2k-2,...2$ and for the energy coefficients we obtain
\begin{align}
w_{r k (\alpha-1)}  & =\sum_{q=0}^{r-1} w_{qk(\alpha+1)} 
+\frac{1}{2}\left(  \left(  \alpha+1\right)  w_{r  k(\alpha+1)} - B_{rk\alpha}\right) ,  \\
\varepsilon_{r  k}  & = w_{r  k1}+2 \sum_{q=0}^{r-1} w_{qk1}.
\end{align}
The solution to $H_{r}$ is given by
\begin{equation}
u_{r}(x,\lambda)= e^{- G_{r}(x,\lambda) } , \qquad \epsilon_{r}(\lambda)=\sum_{k=0}^{\infty}\varepsilon_{rk}\lambda^{k},
\end{equation}
where
\begin{equation}
G_{r}(x,\lambda)=\int w_{r}(x,\lambda)dx = \sum_{k=1}^{\infty}\lambda^{k}\left( \sum_{\alpha=1}^{2k+1}w_{rk\alpha} \frac{x^{\alpha+1}}{\alpha+1} \right).
\end{equation}
Notice that the $\lambda$-independent energy levels in Eq.(\ref{wr0ao}) correspond to the eigenvalues of the harmonic oscillator, thus 
we identify $r$ as the conventional quantum number of the harmonic oscillator. The solution for the only nodeless state 
of $H_{r}$ (the ground state for this Hamiltonian), $u_{r}(x,\lambda)$, provides a new solution to every previously constructed superpartner 
in the set $\{ H_0,H_{1},...,H_{r-1} \}$ which is an excited state and is obtained by the successive action of factorization operators on $u_{r}(x,\lambda)$. 
In particular, for the anharmonic Hamiltonian $H_{0}$  the new unnormalized solution reads
\begin{equation}
u^{0}_{r}(x,\lambda)=a^{\dagger}_{0}a^{\dagger}_{1}...a^{\dagger}_{r-2}a^{\dagger}_{r-1} u_{r}(x,\lambda).
\end{equation} 
We remark that continuing this process we construct the solutions for the infinite set of superpartners $\{ H_0,H_{1},...,H_{r},... \}$. Furthermore,
this solution is valid when we work at a finite order $k$ in the expansion in $\lambda$, thus our procedure yields the solution to an infinite set of Hamiltonians
$\{ H^{(k)}_0,H^{(k)}_{1},...,H^{(k)}_{r},... \}$ at every order $k$.

In summary, the energy for the $r$-th level of the anharmonic oscillator can be written as
\begin{equation}
\epsilon_{r}(\lambda)=\sum_{k=0}^{\infty}\varepsilon_{rk}\lambda^{k},
\end{equation}
while the unnormalized eigenstates are given by
\begin{equation}
u^{0}_{r}(x,\lambda)=a^{\dagger}_{0}a^{\dagger}_{1}...a^{\dagger}_{r-2}a^{\dagger}_{r-1} u_{r}(x,\lambda),
\end{equation}
where 
\begin{align}
u_{r}(x,\lambda)&= e^{-G_{r}(x,\lambda)}, 
\end{align}
is the nodeless solution (ground state) of the $H_{r}$ supersymmetric with
\begin{equation}
G_{r}(x,\lambda)=\int w_{r}(x,\lambda)dx = \sum_{k=1}^{\infty}\lambda^{k}\left( \sum_{\alpha=1}^{2k+1}w_{rk\alpha} \frac{x^{\alpha+1}}{\alpha+1} \right).
\end{equation}

Below we list only the first ten coefficients because of the large expressions 
\begin{align}
\varepsilon_{r0}&=2r+1, \\
\varepsilon_{r1}&=\frac{3}{4}\left(  2r^{2}+2r+1\right), \\
\varepsilon_{r2}&=-\frac{1}{16}\left(  34r^{3}+51r^{2}+59r+21\right), \\
\varepsilon_{r3}&=\frac{1}{64} \left( 375 r^{4}+750 r^{3}+1416 r^{2}+1041r+333\right), \\
\varepsilon_{r4}&=-\frac{1}{1024}\left(  21378r^{5}+53445r^{4}+2\ 71305r^{3}+160470r^{2}+111697r+30885\right),\\
\varepsilon_{r5}&=\frac{1}{4096}\left(  350196r^{6}+1050588r^{5}+3662295r^{4}%
+5573610 r^{3}+6181386 r^{2}+3569679 r+916731\right), \\
\varepsilon_{r6}  & =-\frac{1}{32768}\left(  12529596 r^{7}+43853586 r^{6}%
+190050252 r^{5}+365491665 r^{4}\right.  \nonumber \\
& \left.  +566276728 r^{3}+505850220 r^{2}+277375697 r+65518401\right), \\
\varepsilon_{r7}  & =\frac{1}{131072}\left(  238225977 r^{8}+952903908 r^{7}%
+4961833128 r^{6}+11550335706 r^{5}+23232963558 r^{4}\right. \nonumber \\
& \left.  +28327088832 r^{3}+24506945448 r^{2}+12109639665 r+2723294673\right) , \\
\varepsilon_{r8}  & =-\frac{1}{4194304}\left(  37891923850 r^{9}%
+170513657325 r^{8}+1040570189508 r^{7}+2846265262428 r^{6}\right.  \nonumber \\
& +7092081410526 r^{5}+11012405570670 r^{4}+13234988435964 r^{3}%
+10149533942940 r^{2} \nonumber\\
& \left.  +4834176671621 r+1030495099053\right) ,\\
\varepsilon_{r9}  & =\frac{1}{16777216}\left(  779616467388 r^{10}%
+3898082336940 r^{9}+27355607247375 r^{8}+86033934967860 r^{7}\right.  \nonumber \\
& +257033798474376 r^{6}+486354568850766 r^{5}+756237576702690 r^{4}%
+794460964776060 r^{3}\nonumber \\
& \left.  +594274501768236 r^{2}+264933549728439 r+54626982511455\right), \\
\varepsilon_{r10}  & =-\frac{1}{134217728}\left(  32960936971716 r^{11}%
+181285153344438 r^{10}+1441312485791200 r^{9}+5126267535977115 r^{8}\right. \nonumber \\
& +17927423257122672 r^{7}+40092395638780548 r^{6}+77056659110621118 r^{5}%
+103465523830159170 r^{4} \nonumber\\
& \left.  +107213169962932122 r^{3}+74254844972994534 r^{2}%
+32282806240998167 r+6417007431590595\right).
\end{align}

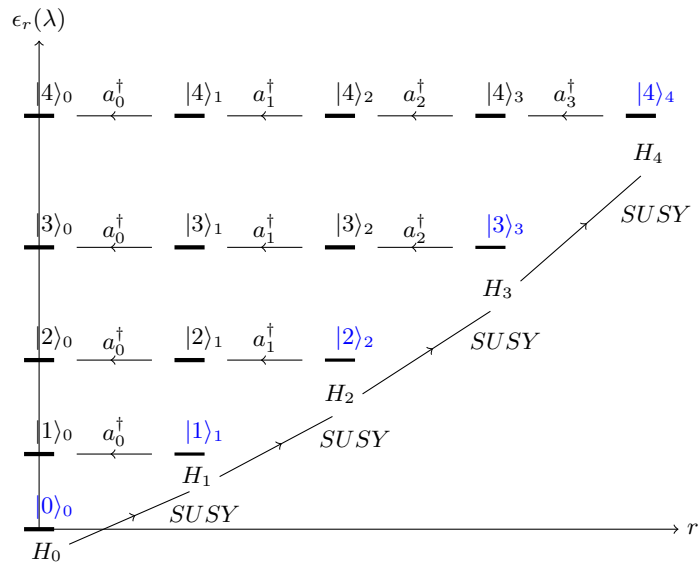
\begin{figure}[ht]
\begin{center}
\begin{tikzpicture}
\draw[black,->] (0, 0)--(0,6.5) node[above]{$\epsilon_{r}(\lambda)$} ;
\draw[black,->] (0, 0)--(8.5,0)  node[right]{$r$} ;
\draw[black,ultra thick] (-0.2,0)--(0.2,0)  node[above ]{{\color{blue}$|0\rangle_{0}$}} ; 
\draw(0.1,-0.3)node{$H_{0}$};
\draw[black,ultra thick] (-0.2,1)--(0.2,1)  node[above ]{$|1\rangle_{0}$} ; 
\draw[black,ultra thick] (-0.2,2.25)--(0.2,2.25)  node[above ]{$|2\rangle_{0}$} ; 
\draw[black,ultra thick] (-0.2,3.75)--(0.2,3.75)  node[above ]{$|3\rangle_{0}$} ; 
\draw[black,ultra thick] (-0.2,5.5)--(0.2,5.5)  node[above ]{$|4\rangle_{0}$} ; 
\draw[fermion] (0.4,-0.2)-- (2,0.5);
\draw(2.2,0.2) node {$SUSY$} ;
\draw[fermion] (1.5,1)-- (0.5,1) ;
\draw(1,1.3) node {$a^{\dagger}_{0}$};
\draw[fermion] (1.5,2.25)-- (0.5,2.25) ;
\draw(1,2.5) node {$a^{\dagger}_{0}$};
\draw[fermion] (1.5,3.75)-- (0.5,3.75) ;
\draw(1,4) node {$a^{\dagger}_{0}$};
\draw[fermion] (1.5,5.5)-- (0.5,5.5) ;
\draw(1,5.8) node {$a^{\dagger}_{0}$};
\draw[black,very thick] (1.8,1)--(2.2,1)  node[above ]{{\color{blue}$|1\rangle_{1}$}} ; 
\draw(2.1,0.7)node{$H_{1}$};
\draw[black,ultra thick] (1.8,2.25)--(2.2,2.25)  node[above ]{$|2\rangle_{1}$} ; 
\draw[black,ultra thick] (1.8,3.75)--(2.2,3.75)  node[above ]{$|3\rangle_{1}$} ; 
\draw[black,ultra thick] (1.8,5.5)--(2.2,5.5)  node[above ]{$|4\rangle_{1}$} ; 
\draw[fermion] (2.4,0.7)-- (3.9,1.5);
\draw(4.2,1.2)node {$SUSY$} ;
\draw[fermion] (3.5,2.25)-- (2.5,2.25) ;
\draw(3,2.55) node {$a^{\dagger}_{1}$};
\draw[fermion] (3.5,3.75)-- (2.5,3.75) ;
\draw(3,4) node {$a^{\dagger}_{1}$};
\draw[fermion] (3.5,5.5)-- (2.5,5.5) ;
\draw(3,5.8) node {$a^{\dagger}_{1}$};
\draw[black,very thick] (3.8,2.25)--(4.2,2.25)  node[above ]{{\color{blue}$|2\rangle_{2}$}} ; 
\draw(4,1.8)node{$H_{2}$};
\draw[black,ultra thick] (3.8,3.75)--(4.2,3.75)  node[above ]{$|3\rangle_{2}$} ; 
\draw[black,ultra thick] (3.8,5.5)--(4.2,5.5)  node[above ]{$|4\rangle_{2}$} ; 
\draw[fermion] (4.3,1.8)-- (6,2.9);
\draw(6.2,2.5)node {$SUSY$} ;
\draw[fermion] (5.5,3.75)-- (4.5,3.75) ;
\draw(5,4) node {$a^{\dagger}_{2}$};
\draw[fermion] (5.5,5.5)-- (4.5,5.5) ;
\draw(5,5.8) node {$a^{\dagger}_{2}$};
\draw[black,very thick] (5.8,3.75)--(6.2,3.75)  node[above ]{{\color{blue}$|3\rangle_{3}$}} ; 
\draw(6.1,3.2)node{$H_{3}$};
\draw[black,ultra thick] (5.8,5.5)--(6.2,5.5)  node[above ]{$|4\rangle_{3}$} ; 
\draw[fermion] (6.4,3.3)-- (8,4.7);
\draw(8.2,4.2)node {$SUSY$} ;
\draw[fermion] (7.5,5.5)-- (6.5,5.5) ;
\draw(7,5.8) node {$a^{\dagger}_{3}$};
\draw[black,ultra thick] (7.8,5.5)--(8.2,5.5)  node[above ]{{\color{blue}$|4\rangle_{4}$}} ; 
\draw(8.1,5)node{$H_{4}$};
\end{tikzpicture}
\end{center}
\caption{Connections between states by the factorization operators of the supersymmetric ladder of Hamiltonians $\{ H_{0},H_{1},H_{2},H_{3},H_{4}\}$ 
required in the calculation of the eigenstates up to $|4\rangle_{0}$ of the anharmonic oscillator Hamiltonian $H_{0}$ using the SEA. The last state to 
right is the edge state of the corresponding Hamiltonian $H_{r}$. }
\label{aosol}
\end{figure}

The scheme of this calculation can be graphically illustrated  similar to Fig.~\ref{aosol} where we use the notation 
$u^{r}_{m}=|r\rangle_{m}$ for the eigenstate corresponding to the $r$ level of the Hamiltonian $H_{m}$ and we show the solution 
up to the $r=4$ level. First we use the logarithmic expansion to solve $H_{0}$ obtaining in this case the solution to the ground state 
$|0\rangle_{0}$ which is the only edge state for the anharmonic potential. Then, we construct the superpartner $H_{1}$ and solve it 
by the same method. The solution, $|1\rangle_{1}$, is the only edge state for $H_{1}$ and turns out to be its ground state. This solution
provides the solution for the first excited state, $|1\rangle_{0}$, of $H_{0}$ obtained by the action of $a^{\dagger}_{0}$ on 
$|1\rangle_{1}$. Next we construct the superpartner $H_{2}$, solve it likewise and from the obtained solution for its unique edge state, 
 $|2\rangle_{2}$, which turns out to be its own ground state, we obtain the first excited solution  $|2\rangle_{1}$  of $H_{1}$ by the action 
 of $a^{\dagger}_{1}$ on $|2\rangle_{2}$ and the second excited state of $H_{0}$ of our interest, $|2\rangle_{0}$, acting with 
 $a^{\dagger}_{0}$ on $|2\rangle_{1}$. The procedure is similar for the calculation of higher excited states of the anharmonic 
 Hamiltonian $H_{0}$. Notice that the supersymmetric structure here is different  to the Hulth\'en or Yukawa case where every 
 superpartner has an edge state for every $n$ level. For the anharmonic case, every superpartner has only one edge state which 
 corresponds to its own ground state as expected from the conventional realization of supersymmetry.

\begin{figure}%
\centering
\includegraphics[width=7cm]{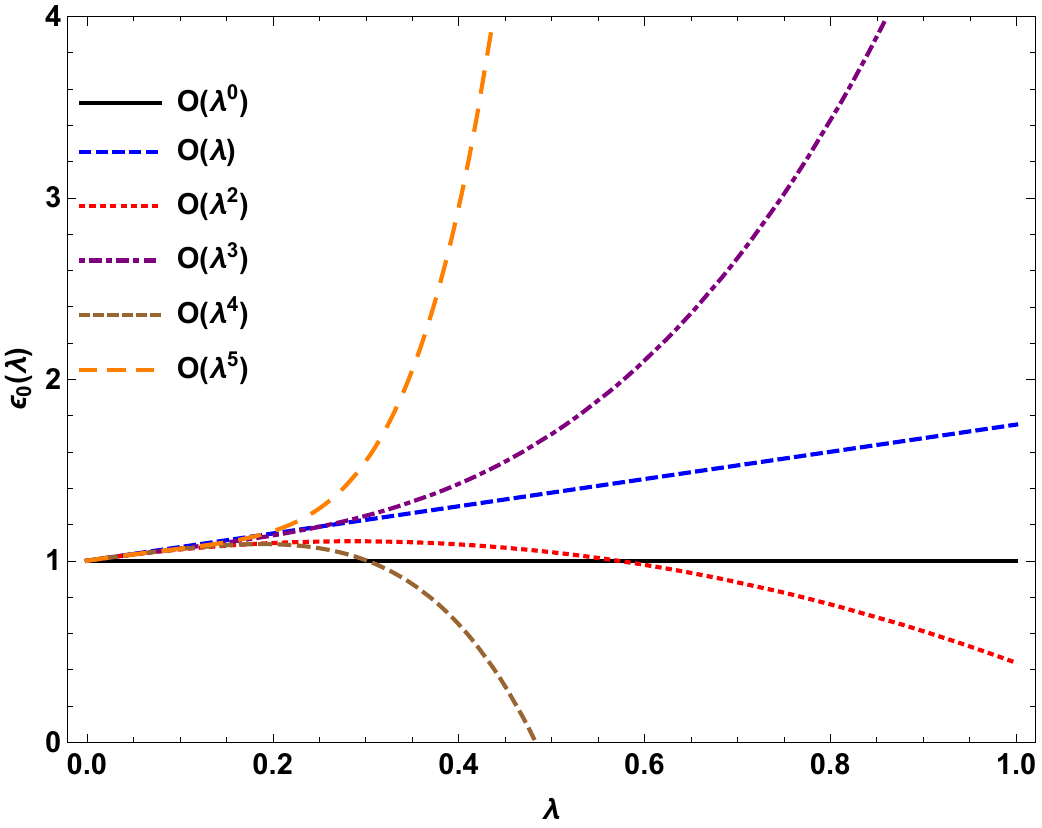} %
\qquad
\includegraphics[width=7cm]{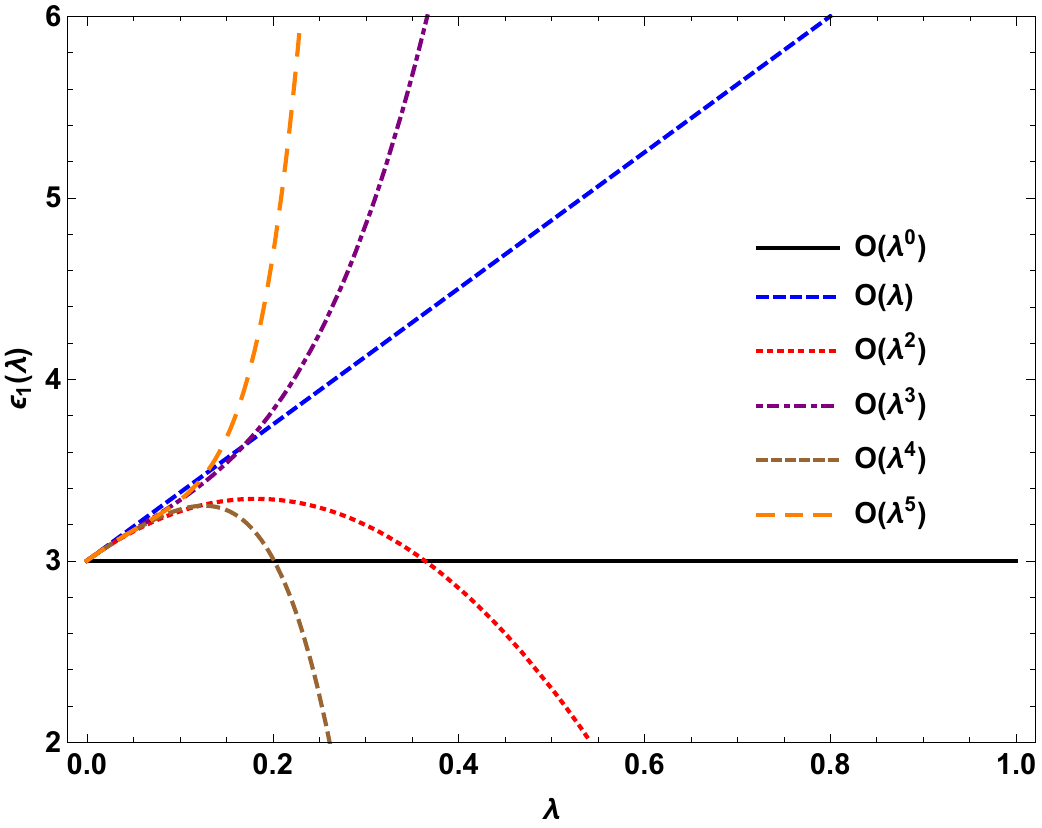} %

\includegraphics[width=7cm]{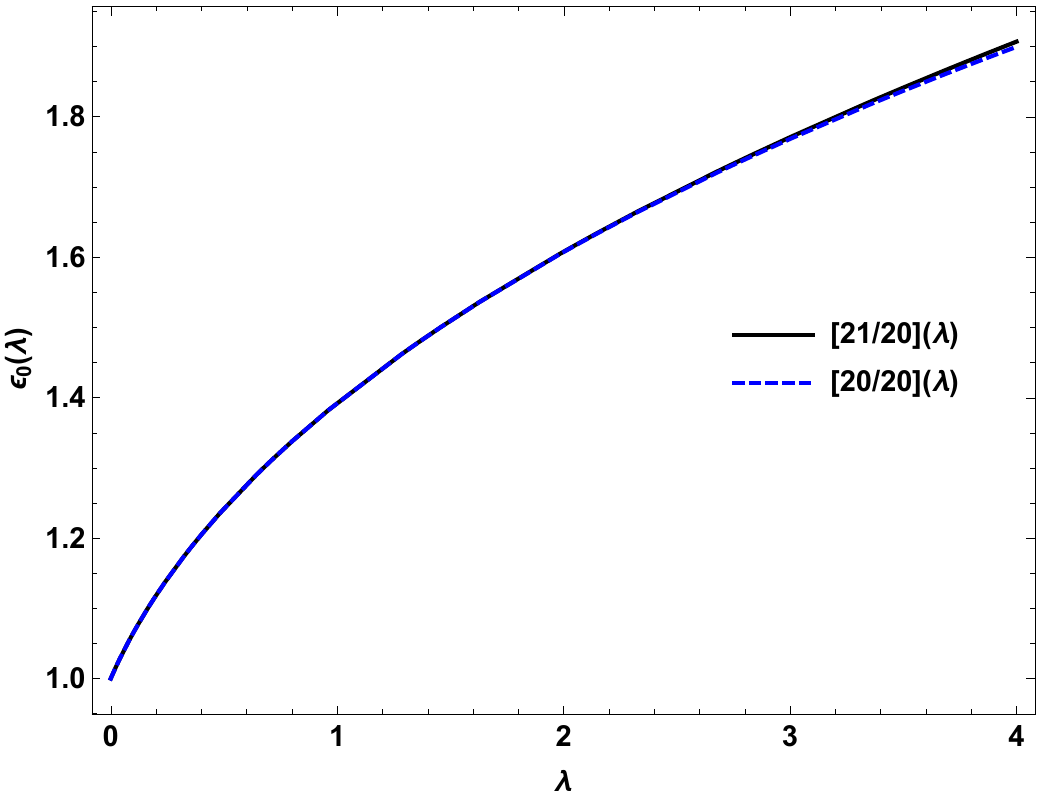} %
\qquad
\includegraphics[width=7cm]{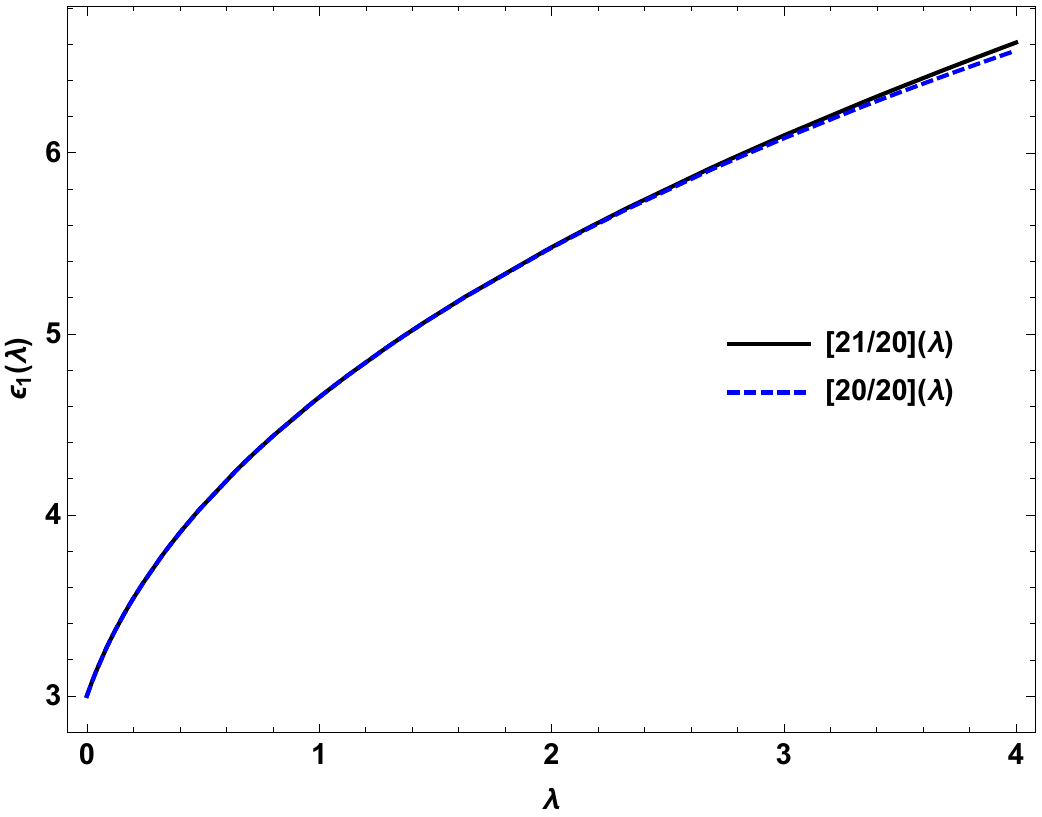} %
\caption{Energy of the ground and first excited states calculated up to order $\lambda^{k}$ for $k=0,1,2,3,4,5$ and reconstruction 
        using the $[21/20](\lambda)$ and $[20/20](\lambda)$ Pad\'{e} approximants.}%
\label{Enhana}%
\end{figure}

As a result of the anharmonic $\lambda x^{4}$ term, the levels are now $\lambda$ dependent and not equally spaced, the separation
of the levels depending on the specific value of $\lambda$. The simplicity of the harmonic oscillator is completely lost and the shape 
invariance property of the harmonic oscillator is lost already at order $\lambda$. 
For the ground energy ($r=0$) of $H_{0}$ the coefficients $\varepsilon_{0k}$ reproduce those of Ref.\cite{Bender:1969si} (the coefficients $A_{k}$ used
in the expansion of the ground energy there are related to our coefficients as $A_{k}=2^{k-1}\varepsilon_{0k}$).  
There are large coefficients in the expressions for $\varepsilon_{rk}$ and the series seems convergent at most for very small values of $\lambda$. In 
Fig. \ref{Enhana} we plot the ground energy and the first excited state calculated to ${\cal{O}}(\lambda^{k})$ for $k=0,1,2,3,4,5$, where 
we can see that the series diverges beyond the very small $\lambda$ region. The functions $\epsilon_{n}(\lambda)$ can be 
reconstructed using techniques like the Pad\'{e} approximants. The figure ~\ref{Enhana} is illustrative of the reconstruction of 
$\epsilon_{0}(\lambda)$ and $\epsilon_{1}(\lambda)$ using the $[21/20](\lambda)$ and $[20/20](\lambda)$ approximation, which requires 
to calculate the series up to ${\cal{O}}(\lambda^{41})$. The actual value 
of the functions lies between these approximants and we can estimate the uncertainty in the reconstruction from the difference between them
for a given $\lambda$ value. Even for values as large as $\lambda=3$, we obtain a precision of one part per thousand in the calculation 
of these energy levels. 

\begin{figure}[ht]
  \begin{center}
  \includegraphics[scale=0.4]{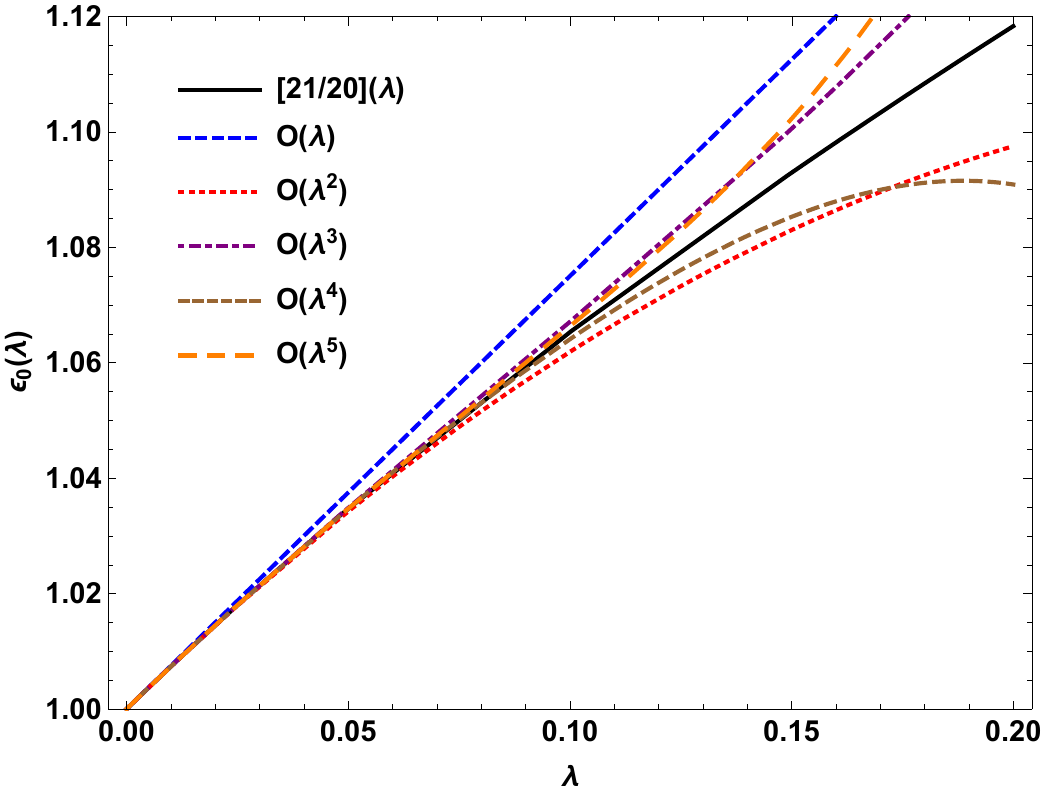}
  \end{center}
  \caption{Ground energy  perturbative calculation up to ${\cal{O}}(\lambda^{5})$ for small $\lambda$ and reconstruction with the 
  $[21/20](\lambda)$ Pad\'{e} approximant.  }
  \label{sm}
\end{figure}

Our results show that because of  the poor convergence of the 
series, care is in order regarding perturbative calculations in anharmonic theories.  The $\lambda\phi^{4}$ potential is used in the standard 
model, where the Higgs coupling has the tree level value 
$\lambda=m^{2}_{H}/2v^{2}$. Using $m_{H}=125$ ~GeV and $v=246$ ~GeV we get $\lambda=0.125$. Our quantum mechanical 
calculations can be considered as calculations in a one-dimensional quantum field theory and in this simple scenario we can check the usefulness 
of the perturbative expansion for this value of $\lambda$. In Fig. \ref{sm} we show the ground energy for small $\lambda$ calculated 
in a perturbative scheme up to 
$\lambda^{5}$, as well as the energy function reconstructed with the $[21/20](\lambda)$ Pad\'{e} approximant. It is clear here, that for 
$\lambda=0.125$ the perturbative calculation up to order $\lambda^{3}$ yields results approaching the actual value of the energy function,
although at  order ${\cal{O}}(\lambda^{4})$ the calculation does not improve that much.  Moreover, at the order of  ${\cal{O}}(\lambda^{5})$ the 
calculation definitively yields the  same precision as the calculation to the order of ${\cal{O}}(\lambda^{3})$. The perturbative expansion 
breaks at ${\cal{O}}(\lambda^{3})$ for $\lambda=0.125$ and higher order terms beyond this point are useful only for the  reconstruction of 
the ground energy using techniques like the Pad\'{e} approximant. This is a one-dimensional field theory result and it 
would be interesting to explore if these results persist for physical four-dimensional theories, which is beyond the scope of this work.

\begin{figure}[ht]
  \begin{center}
  \includegraphics[scale=0.4]{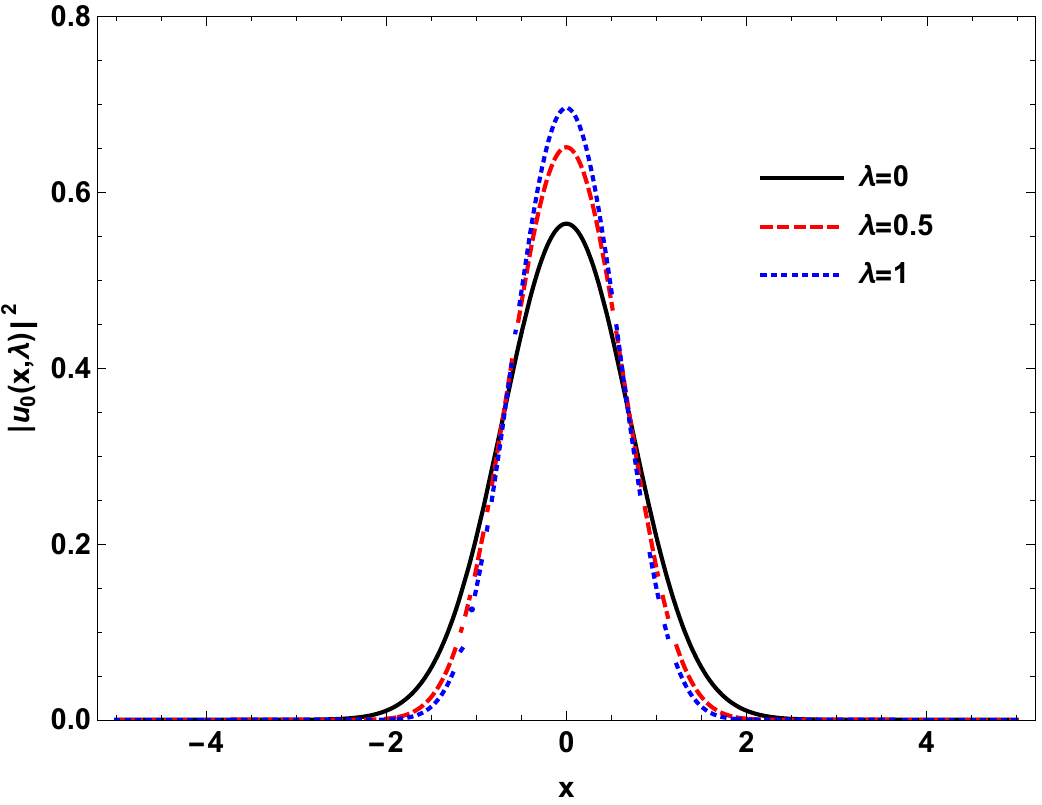}
  \end{center}
  \caption{Probability distribution for the ground state for different values of $\lambda$ reconstructed with the 
  $[6/6](x,\lambda)$ Pad\'{e} approximant.  }
  \label{prob}
\end{figure}

As to the probability distributions, the calculation of the series has also a poor convergence, but calculating enough terms in the series 
we can confidently reconstruct the wave functions for large values of $\lambda$. In Fig.~\ref{prob} we show the probability for the 
ground state as reconstructed with 
the $[6/6](x,\lambda)$ Pad\'{e} approximant. We can see that the shape of this function is similar to the one of  the harmonic oscillator ($\lambda=0$),
and that with the  increase of $\lambda$, the probability becomes  more compact around the origin. Similar results are obtained for the excited states.

\subsection{Supersymmetric perturbation theory}
The case of the anharmonic potential shows that the SEA approach allows to completely solve a problem that has been considered in the past 
using Rayleigh-Schr\"{o}dinger perturbation theory, with  $\lambda x^{4}$ being considered  as a small perturbation of the well known solution 
of the harmonic oscillator. There is nothing technically special in the harmonic oscillator  and we can start with any potential $V(x)$ 
with known solutions and apply the formalism for a general perturbation $\lambda V^{p}(x)$. Indeed, let us consider the general problem
of the potential
\begin{equation}
V(z,\rho,\lambda)= V(z,\rho) +\lambda_{0} V^{p}(z),
\end{equation}
where the solutions to $V(z,\rho)$ are known. Using the length scale associated with $\rho$,  the Schrodinger equation is reduced to
\begin{equation}
\left[-\frac{d^{2}}{dx^{2}} + v(x) + \lambda v^{p}(x)\right]u(\lambda,x)=\epsilon(\lambda)u(\lambda,x).
\end{equation}
We search for the solutions of 
\begin{equation}
H_{0}=-\frac{d^{2}}{dx^{2}} + v_{0}(x),
\end{equation}
where
\begin{equation}
v_{0}(x)= v(x) + \lambda v^{p}(x).
\end{equation}
This potential can be written as 
\begin{equation}
v_{0}(x)=\sum_{k=0}^{\infty}v_{0k}(x)\lambda^{k},
\end{equation}
where
\begin{align}
v_{0k}(x) =\left \{
\begin{array}{r}
v(x) \quad \textit{for} \quad k=0,\\ 
v^{p}(x)\quad \textit{for} \quad k= 1 \\
0 \quad \textit{for} \quad k\ge 2 .
  \end{array}
  \right.
\label{v0kspt}
\end{align}
Following the steps prescribed by the  SEA framework, the cascade of equations for the coefficients $w_{0k}$ in the expansion in $\lambda$ is produced as 
\begin{align}
k&=0:  & w_{00}^{2} - w'_{00} &= v(x) - \varepsilon_{00}, \label{w00spt} \\
k&=1:  & 2 w_{00} w_{01}- w'_{01} &=v^{p}(x) -\varepsilon_{01}, \label{w01spt} \\
k&\ge 2:  & 2 w_{00} w_{0k}- w'_{0k} &= -B_{0k} -\varepsilon_{0k}. \label{w0kspt} 
\end{align}
The solutions to Eq.(\ref{w00spt}) are known and, in cases when the perturbation is a polynomial in $x$ (which covers most of the cases of 
physical interest), we can solve this cascade of equations for the edge states at the desired order, thus  obtaining easily results which have been cumbersome 
to calculate in the conventional Rayleigh-Schr\"{o}dinger perturbation theory. For other states we can construct the set $\{H_{r}\}$ of supersymmetric 
partners or even apply the steps in the SEA to completely solve this problem if necessary. 

\section{Conclusions}
In this work we elaborated a Supersymmetric Expansion Algorithm for obtaining analytical solutions to non-exactly solvable quantum potential problems.
The method incorporates previous achievements in the literature, such as the combination of the logarithmic expansion method and the supersymmetric 
quantum mechanics, next to new elements, such as the concept of  ``edge'' states, to produce a robust framework, applicable to a 
wide class of potentials. Contrary to common belief, we find that for some physical systems of interest (notoriously screened potentials), there exist 
unique excited states which have no nodes. We call them ``edge'' states (excited or ground) and are a new element of the 
No Liouvillian solvable SUSYQM. In the Liouvillian solvable SUSYQM, the ``edge'' states correspond to the ground states of the supersymmetric 
partners in the hierarchy chain. 

The algorithm uses the logarithmic expansion to solve all the edge states of the system and of the auxiliary supersymmetric partners required 
to completely solve the problem. The procedure avoids the old problems with the nodes in the calculation of the excited states, solving first the 
edge states of the supersymmetric set of Hamiltonians and obtaining then the excited states of the system upon the application 
of the factorization operators of supersymmetry on the solutions for the edge states.

The expansions in series of the ingredients of Schr\"odinger's equation, such as energies, wave functions, potentials, and super-potentials, 
produce infinite sets of coupled first order differential equations which can be reduced to hierarchical algebraic 
equations for the coefficients in the expansion and solved exactly order by order, thus providing analytical 
solutions. Along these lines, the three dimensional Hulth\'en potential, and the one dimensional anharmonic oscillator potentials could be completely 
resolved. For Hulth\'en's potential, the solutions present themselves as infinite series in the screening length.  
For zero angular momentum our series reduce to the series expansions of the exact solutions, previously worked out in  \cite{Lam:1971,Flugge:1999}. 
We obtain for the first time  in the literature, power series  solutions for the remaining $\ell\neq 0$ states. The power series solutions are convergent 
for screening lengths below the critical values. We provide precise values for the critical values of the screening for all energy levels 
$\epsilon_{n\ell}$ up to $n=9$ using the power series expansion up to order $\lambda^{30}$ and reconstructing the energy functions with 
Pad\'{e} approximants.    

The power series solutions to the anharmonic potential have a poor convergence, though we could reconstruct the energies as functions of $\lambda$ 
at the desired precision level using Pad\'{e} approximants. We find that even for small  $\lambda$ values, as the value of the quartic 
coupling of the Higgs model in the standard model, $\lambda=0.125$, the perturbative scheme breaks down already at ${\cal{O}}(\lambda^{4})$. 
The eigenstates have a similar form as those of the harmonic oscillator, although they are more compact. The nodes of the excited states are shifted to 
the equilibrium position $x=0$.

These two examples of long pending unsolved potentials,  present just a small illustration  of the predictive power of the Supersymmetric Expansion Algorithm 
which can be  employed for many other purposes, specially as an efficient tool for perturbative calculations.  
It is not the aim of this paper to analyze in  detail the phenomenology of the  numerous  applications the two potentials here considered enjoy 
 in physics and chemistry. We limited ourselves to  generate the solutions which can be used for such purposes and,  for this sake, we will provide 
 freely available codes for the Hulth\'en and anharmonic potentials in the published version of this paper, solving the algebraic recurrence relations 
 which give automatically the solutions to the desired order alongside with some phenomenology of the eigenstates and eigenvalues.


\bibliography{SEA}
\bibliographystyle{JHEP}

\end{document}